\documentclass[runningheads]{llncs}

\usepackage{amsmath}
\usepackage{amsfonts}
\usepackage{bm}
\usepackage{xspace}
\usepackage{graphicx}
\usepackage{array}
\usepackage{rotating}   
\usepackage{multirow}
\usepackage{makecell}
\usepackage{color}
\usepackage{colortbl}
\usepackage{enumerate}
\usepackage{booktabs}
\usepackage{subfig}
\usepackage{wrapfig}
\usepackage{indentfirst}
\usepackage{paralist}
\usepackage{pifont,tabularx,booktabs}
\usepackage{protocolj,arydshln}
\usepackage{array}

\newcolumntype{P}[1]{>{\centering\arraybackslash}p{#1}}

\newcommand{\Hzk}{\ensuremath{\mathsf{H}_{\mathsf{zk}}}}

\newcommand{\name}{BBB-Voting\xspace}

\usepackage[normalem]{ulem}
\newcommand{\cmark}{\textcolor{black}{\ding{51}}}
\newcommand{\xmark}{\textcolor{black}{\ding{55}}}
\newcommand{\trot}[1]{\multicolumn{1}{l}{\rlap{\rotatebox{25}{#1}~}}}

\newcommand{\specialcell}[2][c]{
	\begin{tabular}[#1]{@{}l@{}}#2\end{tabular}}

\usepackage{hyperref} 
\usepackage[nameinlink,capitalize]{cleveref}

\renewcommand{\tablename}{\color{black}Table}
\renewcommand{\figurename}{\color{black}Figure}

\newcommand{\myparagraph}[1]{\vspace{0.1cm}\noindent{\it #1.}}

\let\llncssubparagraph\subparagraph
\let\subparagraph\paragraph
\usepackage[compact]{titlesec}
\let\subparagraph\llncssubparagraph
\usepackage[inline]{enumitem}
\setlist[enumerate,1]{label=\textit{\alph*)}}
\usepackage[font={small,it},skip=3pt,belowskip=-7pt]{caption}

\renewcommand{\tablename}{Tab.}
\renewcommand{\figurename}{Fig.}

\usepackage{listings}
\definecolor{backcolour}{rgb}{0.95,0.95,0.92}
\lstdefinestyle{lststyle}{
	basicstyle=\scriptsize,
	breakatwhitespace=false,
	breaklines=true,
	captionpos=b,
	keepspaces=true,
	numbersep=5pt,
	showspaces=false,
	showstringspaces=false,
	showtabs=false,
	tabsize=2
}

\makeatletter
\g@addto@macro\normalsize{
	\setlength\abovedisplayskip{1.1pt}
	\setlength\belowdisplayskip{1.1pt}
	\setlength\abovedisplayshortskip{1.1pt}
	\setlength\belowdisplayshortskip{1.1pt}
}

\AtBeginDocument{
	}

\begin{document}
\title{BBB-Voting: Self-Tallying End-to-End
	Verifiable 1-out-of-$k$ Blockchain-Based Boardroom Voting \vspace{-0.5cm}}

\author{Sarad Venugopalan\inst{1}$^\ddagger$ \and
	Ivan Homoliak\inst{2}$^\ddagger$ \and
	Zengpeng Li\inst{1} \and 
	Pawel Szalachowski\inst{1}
}
\authorrunning{S. Venugopalan et al.}
\institute{Singapore University of Technology and Design \and
	Brno University of Technology, Faculty of Information Technology \\ $^\ddagger$\textit{The authors contributed equally.}}

\maketitle              

\vspace{-0.8cm}
\begin{abstract}
Voting is a means to agree on a collective decision based on available choices (e.g., candidates), where participants agree to abide by their outcome. 
To improve some features of e-voting, decentralized blockchain-based solutions can be employed, where the blockchain represents a public bulletin board that in contrast to a centralized bulletin board provides extremely high availability, censorship resistance, and correct code execution.
A blockchain ensures that all entities in the voting system have the same view of the actions made by others due to its immutability and append-only features. 
The existing remote blockchain-based boardroom voting solution called Open Voting Network (OVN) provides the privacy of votes, universal \& End-to-End verifiability, and perfect ballot secrecy; however, it supports only two choices and lacks robustness enabling recovery from stalling participants. 

We present BBB-Voting, an equivalent blockchain-based approach for decentralized voting such as OVN, but in contrast to it, BBB-Voting supports 1-out-of-$k$ choices and provides robustness that enables recovery from stalling participants. 
We make a  cost-optimized implementation using an Ethereum-based environment respecting Ethereum Enterprise Alliance standards, which we compare with OVN and show that our work decreases the costs for voters by $13.5\%$ in normalized gas consumption.
Finally, we show how BBB-Voting can be extended to support the number of participants limited only by the expenses paid by the authority and the computing power to obtain the tally.

\vspace*{-0.3cm}
\keywords{Electronic Voting  $\bullet$ Blockchain $\bullet$ End-to-End Verifiability $\bullet$ Self-Tallying$\bullet$ Robustness $\bullet$ Scalability $\bullet$ Ethereum Enterprise Alliance.}

\end{abstract}

\section{Introduction}
\label{sec:introduction}
Voting is an integral part of democratic governance, where eligible participants can cast a vote for their representative choice (e.g., candidate or policy) through a secret ballot.
The outcome of voting announces a tally of votes.
Voting is usually centralized and suffers from a single point of failure that can be manifested in censorship, tampering, and issues with availability of a service.

Blockchain is an emerging decentralized technology that provides interesting properties such as decentralization, censorship-resistance, immutability of data, correct execution of code, and extremely high availability, which can be harnessed in addressing the existing issues of e-voting.
A few blockchain-based e-voting solutions have been proposed in recent years, mostly focusing on boardroom voting~\cite{McCorrySH17,EPRINT:PanRoy18,Li2020,yu2018platform} or small-scale voting~\cite{DBLP:conf/fc/SeifelnasrGY20,icissp:DMMM18,Li2020}.

Decentralization was a desired property of e-voting even before invention of blockchains.
For example, (partially) decentralized e-voting that uses the homomorphic properties of El-Gamal encryption was introduced by Cramer et al.~\cite{cgs97}. 
It assumes a threshold number of honest election authorities to provide the privacy of vote.
However, when this threshold is adversarial, it does not protect from computing partial tallies, making statistical inferences about it, or even worse  -- the vote choices of participants.
A solution that removed trust in tallying authorities was for the first time proposed by Kiayias and Yung~\cite{Kiayias2002} in their privacy-preserving self-tallying boardroom voting protocol.
A similar protocol was later proposed by Hao et al.~\cite{HaoRZ10}, which was later extended to a blockchain environment by McCorry et al.~\cite{McCorrySH17} in their Open Voting Network (OVN).
An interesting property of OVN is that it requires only a single honest voting participant to maintain the privacy of the votes.
However, OVN  supports only two vote choices (based on~\cite{HaoRZ10}), assumes no stalling participants, and requires expensive on-chain tally computation (limiting its scalability).
The scalability of OVN was improved by Seifelnasr et al.~\cite{DBLP:conf/fc/SeifelnasrGY20}, but retaining the limitation of 2 choices and missing robustness.

Our goal is to build a remote boardroom voting protocol that resolves these limitations and would enable a straightforward extension to support scalability.
Therefore, we introduce BBB-Voting, a blockchain-based boardroom voting system providing  1-out-of-$k$ voting, while additionally offering a mechanism for resolution of faulty participants.
Alike OVN, BBB-Voting also provides the maximum privacy of votes in the setting that outputs the full tally of votes (as opposed to \textit{tally-hiding} protocols~\cite{kusters2020ordinos,huber2022kryvos}). 
Both OVN and BBB-Voting require the authority
whose role is limited to registering participants into voting and shifting the phases of the protocol (also possible by the participants).
The communication between the participants and the blockchain is semi-synchronous; i.e.,
each participant is expected to execute certain actions within a given time frame. 
When all registered participants submit their votes, the result can be tallied by anybody and the correctness of the result is verified by the blockchain.

\myparagraph{\textbf{Contributions}}
We make the following contributions.
\begin{compactenum}[i)]
	\item We present \name, an approach for remote end-to-end verifiable priva\-cy-preserving self-tallying 1-out-of-$k$ boardroom voting on the blockchain (see \autoref{sec:protocol}).	
	In detail, we start with the voting protocol proposed by Hao et al.~\cite{HaoRZ10} that provides a low bandwidth requirements and computational costs but is limited to 2 vote choices.
	We extend this protocol to support $k$ choices utilizing the 1-out-of-$k$ proof verification proposed by Kiayias and Yung~\cite{Kiayias2002}. 
	We accommodate this approach to run on the blockchain with Turing-complete smart contract capability, enabling on-chain zero-knowledge proof verification of blinded votes (and other proofs).

	\item We incorporate a robustness approach~\cite{KhaderSRH12} into our protocol, which enables us to eliminate (even reoccurring) stalling (i.e., faulty) participants and thus finish voting without restarting the protocol (see \autoref{sec:faulttolerance}).
	
	\item We make two implementations, one based on discrete logarithm problem (DLP) for integers modulo $p$ and the second one based on the elliptic curve DLP.
	For both implementations we propose various optimizations reducing the costs imposed by the blockchain platform (\autoref{ssec:optimizatations}).
	Due to the optimizations, our implementation (with elliptic curve DLP) increases the number of participants fitting a single block by $9\%$ in contrast to OVN~\cite{McCorrySH17} under the same assumptions, while it decreases the costs for voters by $13.5\%$.

	\item We outline a scalability extension of our work, enabling the number of participants to be limited only by the expenses paid by the authority to register participants and compute their MPC keys as well as the computing power to obtain the tally.
	For demonstration purposes, we evaluate its utility in the context of the voting that is a magnitude greater than the boardroom voting (i.e., up to 1000 participants) while preserving almost the same per-participant costs paid by the authority as without this extension (see \autoref{ssec:scalability-improvements}).

\end{compactenum}

\section{Preliminaries}
\label{sec:preliminaries}

In this section, we describe terminology specific to voting. 
We assume that the reader is familiar with blockchains and smart contracts, and we refer an unfamiliar reader to \autoref{sec:appendix-blockchains} for a brief background.

\smallskip
An \textit{involved party} refers to any stakeholder of the voting process and it covers \textit{all participants} and the \textit{authority}. 
A voting protocol is expected to meet several properties.
A list of such properties appears in the works of Kiayias and Yung~\cite{Kiayias2002}, Groth~\cite{Groth2004}, and Cramer et al.~\cite{cgs97}. 

		\noindent\textbf{(1) Privacy of the Vote:} 
		ensures the secrecy of the ballot contents~\cite{Kiayias2002}.
		Hence, in a scheme that meets this property, a participant's vote must not be revealed other than by the participant herself upon her discretion (or through the collusion of all remaining participants).
		Usually, privacy is ensured by trusting authorities in traditional elections or by homomorphic encryption in some decentralized e-voting solutions (e.g.,~\cite{Kiayias2002,HaoRZ10,McCorrySH17,DBLP:conf/fc/SeifelnasrGY20,icissp:DMMM18}). 
		
		\noindent\textbf{(2) Perfect Ballot Secrecy:} is an extension of the privacy of the vote. 
		It implies that a partial tally (i.e., prior to the end of voting) is available only if all remaining participants are involved in its computation.	
				
		\noindent\textbf{(3) Fairness:} ensures that a tally may be calculated only after all participants have submitted their votes.
		Therefore, no partial tally can be revealed to anyone before the end of the voting protocol~\cite{Kiayias2002}.
		
		\noindent\textbf{(4a) Universal Verifiability:} any involved party can verify that all cast votes are correct and they are correctly included in the final tally~\cite{Kiayias2002}.
		
		\noindent\textbf{(4b) End-to-End (E2E) Verifiability:}
		The verifiability of voting systems is also assessed by E2E verifiability~\cite{benaloh2015end}, which involves \textit{cast-as-intended}, \textit{recorded-as-cast}, and \textit{tallied-as-recorded} verifiability~\cite{jonker2013privacy}.

		\noindent\textbf{(5) Dispute-Freeness:} extends the notion of verifiability.
		A dispute-free~\cite{Kiayias2002} voting protocol contains built-in mechanisms eliminating disputes between participants; therefore, anyone can verify whether a participant followed the protocol. 
		Such a scheme has a publicly-verifiable audit trail that contributes to the reliability and trustworthiness of the scheme.

		\noindent\textbf{(6) Self-Tallying:} once all the votes are cast, any involved party can compute the tally.
		Self-tallying systems need to deal with the fairness issues (see (3) above) because the last participant is able to compute the tally even before casting her vote.
		This can be rectified with an additional verifiable dummy vote~\cite{Kiayias2002}.

		\noindent\textbf{(7) Robustness (Fault Tolerance):} the voting protocol is able to recover from faulty (stalling) participants, where faults are publicly visible and verifiable due to dispute-freeness~\cite{Kiayias2002}. 		
		Fault recovery is possible when all the remaining honest participants are involved in the recovery.

		\noindent\textbf{(8) Resistance to Serious Failures:}
		 Serious failures are defined as situations in which voting results were changed either by a simple error or an adversarial attack. 
		 Such a change may or may not be detected. 
		 If detected in non-resistant systems, it is irreparable without restarting the entire voting~\cite{park2021going}.

\section{System Model \& Overview}
\label{sec:model}

\iffalse
\begin{figure}
	\centering
	\includegraphics[width=1\columnwidth]{fig/booth.pdf}
	\caption{Polling booths are replicated instances of a smart contract. Each contract instance corresponds to a single booth and serves a limited number of participants. New instances (booths) are added on demand to provide scalability.}
	\label{fig:booth}
\end{figure}
\fi

Our system model has the following actors and components:
(1) \textit{a participant (P)}, who votes for a choice of her decision,
(2) \textit{a voting authority ($VA$)} who is responsible for validating the eligibility of $P$s to vote, their registration, and 
(3) \textit{a smart contract (SC)}, which collects the votes, acts as a verifier of zero-knowledge proofs, enforces the rules of voting, and verifies the tally.

\subsection{Adversary Model}\label{sec:adversary-model}
The adversary $\mathcal{A}$ has bounded computing power, is unable to break used cryptographic primitives, and can control at most $t$ of $n$ participants during the protocol, where $t\leq n-2$ $\wedge$ $n\geq3$ (see \autoref{sec:appendixvoteorivacy}).
Any $P$ under the control of $\mathcal{A}$ can misbehave during the protocol execution. 
$\mathcal{A}$ is also a passive listener of all communication entering the blockchain network but cannot block it or replace it with a malicious message since all transactions sent to the blockchain are authenticated by signatures of $P$s or $VA$.
Finally, we assume that $VA$ \textbf{is only trusted in terms of identity management}, i.e., it performs identity verification of $P$s honestly, and neither censor any $P$ nor register any spoofed $P$.
Nevertheless, no trust in $VA$ is required to execute the rest of the protocol.

\section{BBB-Voting Scheme}
\label{sec:protocol}

\name scheme provides all properties mentioned in \autoref{sec:preliminaries}.
Similar to OVN~\cite{McCorrySH17}, \name publishes the full tally at the output and uses homomorphic encryption to achieve privacy of votes and perfect ballot secrecy.
In detail, we extend the protocol of Hao et al.~\cite{HaoRZ10} to support $k$ choices utilizing the 1-out-of-$k$ proof verification proposed by Kiayias and Yung~\cite{Kiayias2002}, and we accommodate this approach to run on the blockchain. 
Additionally, we extend our protocol to support the robustness, based on Khader et al.~\cite{KhaderSRH12},
which enables the protocol to recover (without a restart) from a number of faulty participants who did not submit their votes.\footnote{This number depends on the blockchain platform's block gas limit and in our evaluation that sets 10M for block gas limit, it is equal to 9 participants.}
As a consequence, robustness increases the resistance of our protocol to serious failures.

\subsection{Base Variant}
\label{ssec:basicprotocol}
\begin{wrapfigure}{t}{0.55\textwidth}
	\centering
	\vspace*{-0.1cm}
	
	\includegraphics[width=\linewidth]{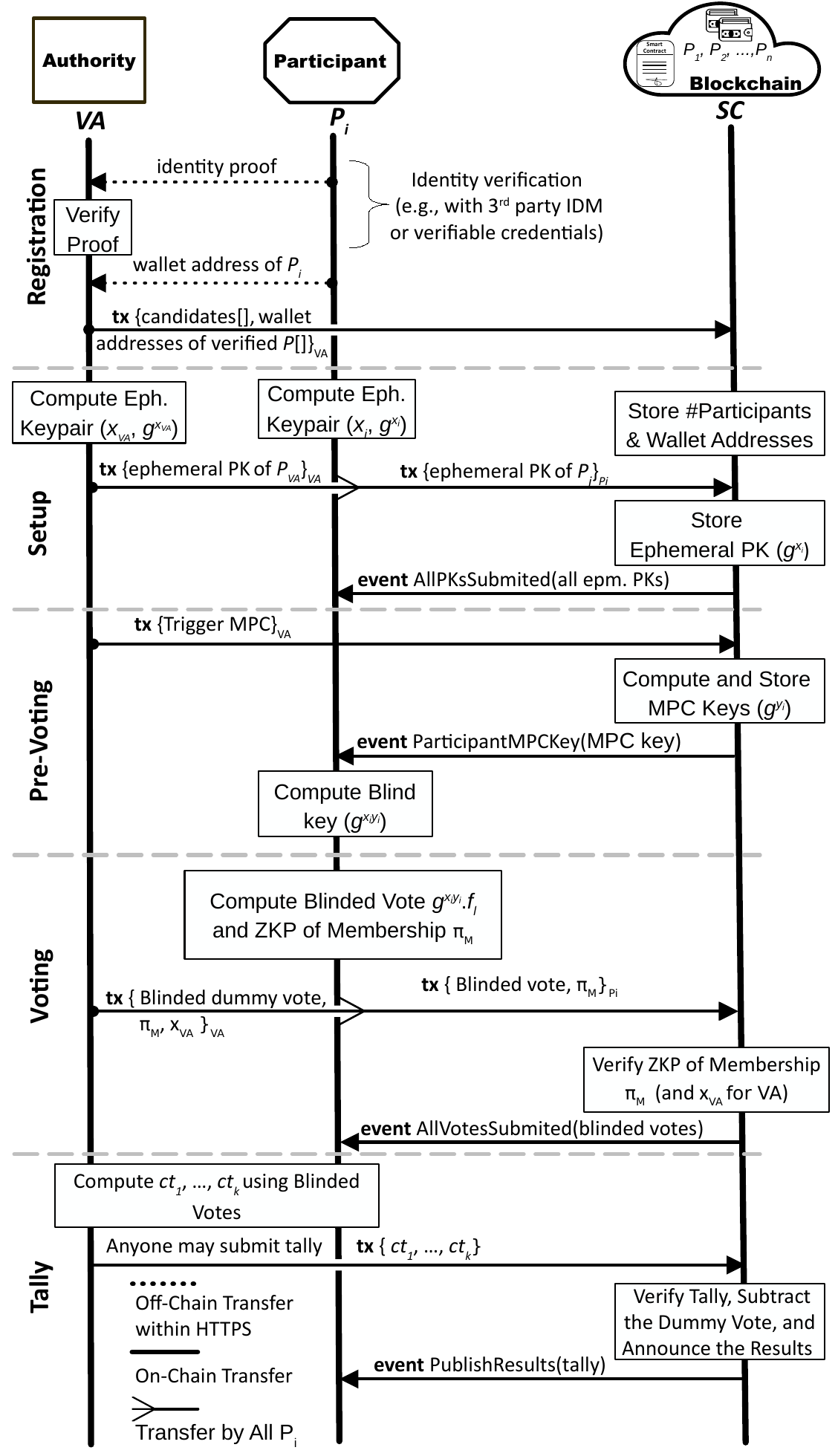}
	\caption{Basic protocol of \name. The notation $tx(param)_{\mathbb{A}}$ corresponds to a transaction, digitally signed by $\mathbb{A}$'s wallet private key.
	}
	\label{fig:operation-scheme-bw}
	\vspace*{-0.8cm}
\end{wrapfigure}
key ownership verification, enrollment at $SC$), a \textbf{setup} (an agreement on system parameters, submission of ephemeral public keys), \textbf{pre-voting} 
The base variant of \name does not involve a fault recovery and is divided into five stages, each describing the actions taken by $P$s and $VA$ in a distributed setting (see \autoref{fig:operation-scheme-bw}).
It involves a \textbf{registration} (identity verification, (computation of MPC keys), \textbf{voting} (vote packing, blinding, and verification), and \textbf{tally} phases.
All faulty behaviors of $P$s and $VA$ are subject to deposit-based penalties. 
In detail, $P$ who submitted her ephemeral key (in the setup phase) and then has not voted within the timeout will lose the deposit. 
To achieve fairness, $VA$ acts as the last $P$ who submits a ``dummy'' vote with her ephemeral private key\footnote{Privacy for a dummy vote is not guaranteed since it is subtracted from the tally.} after all other $P$s cast their vote (or upon the voting timeout expiration).

\subsubsection{\textbf{Phase 1 (Registration)}}
$VA$ first verifies the identity proof of each $P$.
For decentralized identity management (IDM), the identity proof is represented by the verifiable credentials (VC)~\cite{verifiable-credentails} signed by the issuer, while in a centralized IDM the identity proof is interactively provided by a third-party identity provider (e.g., Google).
First, $VA$ verifies the issuer's signature on the identity proof. 
Next, $VA$ challenges $P$ to prove (using her VC) that she is indeed the owner of the identity.
Further, each $P$ creates her blockchain wallet address (i.e., the blockchain public key (PK)) and provides it to $VA$.
The $VA$ locally stores a bijective mapping between a $P$'s identity and her wallet address.\footnote{Note that the address of $P$ must not be part of identity proof to preclude $VA$ from owning an indisputable proof of identity to blockchain address mapping (see \autoref{sec:blockchain-specific-issues}).} 
Next, $VA$ enrolls all verified $P$s by sending their wallet addresses to $SC$.

\subsubsection{\textbf{Phase 2 (Setup)}}
	$P$s agree on system parameters that are universal to voting -- the parameters for voting are publicly visible on $SC$ (deployed by $VA$ in a transaction).
	Therefore, any $P$ may verify these parameters before joining the protocol.
	Note the deployment transaction also contains the specification of timeouts for all further phases of the protocol as well as deposit-based penalties for misbehavior of $VA$ and $P$s.
	The parameters for voting are set as follows:

\setlength{\parskip}{5pt} \setlength{\itemsep}{5pt}
\begin{compactenum}[$\;$]
	\item[1)] $VA$ selects a common generator $g\in \mathbb{F}_p^*$. 
	The value of $p$ is chosen to be a safe prime, i.e., $p=2\cdot  q +1$, where $q$ is a prime. 
	A safe prime is chosen to ensure the multiplicative group of order  $p - 1 = 2\cdot q$, which has no small subgroups that are trivial to detect.\footnote{We use modular exponentiation by repeated squaring to compute $g^x$ $mod$ $p$, which has a time complexity of $\mathcal{O}$(($log$ $x$)$\cdot (log^2$ $p$))~\cite{Koblitz}. } 
	Let $n < p - 1$.	
	\item[2)] Any participant $P_i$ is later permitted to submit a vote $\{v_i ~|~ i\in \{1,2,...,k\}\}$ for one of $k$ choices. This is achieved by selecting $k$ independent generators $\{f_1,...,f_k\}$  in $\mathbb{F}_p^*$ (one for each choice).
	These generators for choices should meet a property described by Hao et al.~\cite{HaoRZ10} to preclude having two different valid tallies that fit \autoref{eqn:eq3}:
	\begin{eqnarray}
		f_i = 
		\begin{cases}
		g^{2^0}       & \quad \text{for choice 1},\\
		g^{2^m}       & \quad \text{for choice 2},\\
		\quad \quad \quad \cdots \\
		g^{2^{(k-1)m}} & \quad \text{for choice k},
		\end{cases}
	\end{eqnarray}
	where $m$ is the smallest integer such that $2^m > n$ (the number of participants).		
\end{compactenum}

	\noindent\textbf{Ephemeral Key Generation \& Committing to Vote.}
	Each $P_i$ creates her ephemeral private key as a random number $x_i\in_{R} \mathbb{F}_p^*$ and ephemeral public key as $g^{x_i}$.
	Each $P_i$ sends her ephemeral public key to $SC$ in a transaction signed by her wallet, thereby, committing to submit a vote later.\footnote{Note that in contrast to OVN~\cite{McCorrySH17} (based on the idea from~\cite{HaoRZ10}), we do not require $P_i$ to submit ZKP of knowledge of $x_i$ to $SC$ since $P_i$ may only lose by submitting $g^{x_i}$ to which she does not know $x_i$ (i.e., losing a chance to vote and the deposit). 
	}
	Furthermore, $P_i$ sends a deposit in this transaction, which can be retrieved back after the end of voting.
	However, if $P_i$ does not vote within a timeout (or does not participate in a fault recovery (see \autoref{sec:faulttolerance})), the deposit is lost, and it is split to the remaining involved parties.
	$P$s who do not submit their ephemeral keys in this stage are indicating that they do not intend to vote; the protocol continues without them and they are not subject to penalties.		
	Finally, each $P$ obtains (from $SC$) the ephemeral public keys of all other verified $P$s who have committed to voting.
	Note that ephemeral keys are one-time keys, and thus can be used only within one run of the protocol to ensure privacy of votes (other runs require fresh ephemeral keys).

\subsubsection{\textbf{Phase 3 (Pre-Voting)}}
This phase represents multiparty computation (MPC), which is run to synchronize the keys among all $P$s and achieve the \textit{self-tallying} property.
However, no direct interaction among $P$s is required since all ephemeral public keys are  published at $SC$. The MPC keys are computed by $SC$, when $VA$ triggers the compute operation via a transaction.
The $SC$ computes and stores the MPC key for each $P$ as follows:
	\begin{equation}
	\label{eqn:eq1}
	h=g^{y_i}=\prod\limits_{j=1}^{i-1} g^{x_j}/\prod\limits_{j=i+1}^{n} g^{x_j},
	\end{equation}
	where $y_i=\sum_{j<i}x_j- \sum_{j>i}x_j$
	and
	$\sum_{i}x_i y_i=0$ (see Hao et al.~\cite{HaoRZ10} for the proof).
	Since $y_i$ is derived from the private ephemeral keys of all other $P$s it is secret. 
	While anyone can compute $g^{y_i}$, to reveal $y_i$, all $P$s $\setminus $ $P_i$ must either collude or solve the DLP for \autoref{eqn:eq1}.\footnote{Note that this DLP was already computed for 795-bit long safe prime in 2019~\cite{integer-dlp-2019}; the computation took $\sim3100$ core-years, using Intel Xeon Gold 6130 (2.1GHz) CPUs.} 
	As the corollary of \autoref{eqn:eq1}, the protocol preserves vote privacy if there are at least 3 $P$s with at least 2 honest (see the proof in \autoref{sec:appendixvoteorivacy}).

\begin{figure}[t]
	\begin{center}
		\scriptsize
		\vspace{-0.3cm}
		\fbox{
			\begin{protocolm}{2}						
				\participants{\underline{Participant $P_i$}}{\underline{Smart Contract}}
				\participants{($~h\leftarrow g^{y_i},~ v_i$)}{($~h\leftarrow g^{y_i}$)}
				\hline	
				& &\\		
				Select~v_i\in\{1,...,k\}, & & \\	 				
				Use~choice~generators~\\
				f_l\in\{f_1,...,f_k\}\subseteq\mathbb{F}_p^*,\\			
				Publish~ x\leftarrow g^{x_i} && 	\\
				Publish~ B_i\leftarrow h^{x_i}{f_l} & & \\

				w\in_{R} \mathbb{F}_p^*\\

				\forall l\in \{1,..,k\}\setminus {v_i}:\\
				
				\quad 1.~r_l,d_l\in_{R} \mathbb{F}_p^* \\
				\quad 2.~a_l\leftarrow x^{-d_l}g^{r_l}\\ 				
				\quad 3.~b_l\leftarrow h^{r_l}(\frac{B_i}{f_l})^{-d_l} \\

				for\enspace {v_i}:\\
				
				\quad 1.~a_{v_i}\leftarrow g^{w}\\
				\quad 2.~b_{v_i}\leftarrow h^{w} \\
				\hdashline
				c\gets\Hzk(\{ \{a_l\},\{b_l\} \}_{l})\\
				\hdashline
				
				for\enspace {v_i}:\\
				
				\quad 1.~d_{v_i}\leftarrow \sum_{l\neq{v_i}}d_l\\
				\quad 2.~d_{v_i}\leftarrow c - d_{v_i}\\
				
				\quad 3.~r_{v_i}\leftarrow w+x_id_{v_i} \\			
				\quad 4.~q\leftarrow p-1\\
				\quad 5.~{r_{v_i}\leftarrow r_{v_i}\mod{q}}\\

				&\sends{\forall l: \{a_l\}, \{b_l\}, \{r_l\},\{d_l\}}
				&\\
				
				& &\Psi \gets \{\forall l: \{a_l\},\{b_l\}\}  \\
				&& c \gets\Hzk(\Psi)\\
				
				& &\sum_{l} d_{l} \stackrel{?}{=} c\\
				
				&& \forall l\in \{1,..,k\}\\

				&&\quad 1.~g^{r_l}\stackrel{?}{=}{a_l}x^{d_l}\\
				&&\quad 2.~h^{r_l}\stackrel{?}{=}{b_l}(\frac{B_i}{{f_l}})^{d_l}\\
			\end{protocolm}
		}
	\end{center}
	\vspace{-0.35cm}
	\caption{Zero knowledge proof of set membership for 1-out-of-$k$ choices. }
	\label{fig:ZKPMultiCandidate}
	\vspace{-0.5cm}
\end{figure}	

\subsubsection{\textbf{Phase 4 (Voting)}}
In this phase, each $P_i$  blinds and submits her
vote to $SC$.
These steps must ensure the recoverability of the tally, vote privacy, and well-formedness of the vote.
Vote privacy is achieved by multiplying the $P_i$'s blinding key with her vote choice.
The blinded vote of the participant $P_i$ is
\begin{equation}
\label{eqn:eq2}
B_i=\begin{cases}
g^{x_iy_i}f_1 & \text{if $P_i$ votes for choice 1},\\
g^{x_iy_i}f_2 & \text{if $P_i$ votes for choice 2},\\
~~~~~~~~~~~~~...\\
g^{x_iy_i}f_k & \text{if $P_i$ votes for choice \textit{k}.}
\end{cases}
\end{equation}
The $P$ sends her choice
 within a blinded vote along with a  1-out-of-$k$ non-interactive zero-knowledge (NIZK) proof of set membership to $SC$.
We modified the approach proposed in Kiayias et al.~\cite{Kiayias2002} to the form used by Hao et al.~\cite{HaoRZ10}, which is convenient for practical deployment on existing smart contract platforms.
The verification of set membership using this protocol is depicted in \autoref{fig:ZKPMultiCandidate}, where $P_i$ is a prover and $SC$ is the verifier.
Hence, $SC$ verifies the correctness of the proof and then stores the blinded vote.
In this stage, it is important to ensure that no re-voting is possible, which is to avoid any inference about the final vote of $P$ in the case she would change her vote choice during the voting stage.
Such a re-voting logic can be enforced by $SC$, while user interface of the $P$ should also not allow re-voting.
Moreover, to ensure fairness, $VA$ acts as the last $P$ who submits a dummy vote together with her ephemeral private key.

\subsubsection{\textbf{Phase~5~(Tally)}}
When the voting finishes (i.e., voting timeout expires or all $P$s and $VA$ cast their votes), the tally of votes received for each of $k$ choices is computed off-chain by any party and then submitted to $SC$.
When $SC$ receives the tally, it verifies whether \autoref{eqn:eq3} holds, subtracts a dummy vote of $VA$, and notifies all $P$s about the result. 
The tally is represented by vote counts $ct_i,\forall i\in\{1,...,k\}$ of each choice, which are computed using an exhaustive search fitting 
\begin{equation}
\label{eqn:eq3}
\prod\limits_{i=1}^{n}B_i=
\prod\limits_{i=1}^{n}g^{x_iy_i}f=
g^{\sum_{i}x_i y_i}f=
{f_1}^{ct_1}{f_2}^{ct_2}...{f_k}^{ct_k}.
\end{equation}
The maximum number of attempts is bounded by combinations with repetitions to ${n+k-1}\choose{k}$.
Although the exhaustive search of 1-out-of-$k$ voting is more computationally demanding in contrast to 1-out-of-2 voting~\cite{McCorrySH17},~\cite{HaoRZ10}, this process can be heavily parallelized. 
See time measurements in \autoref{sec:tally-exps}.

\subsection{Variant with Robustness}
\label{sec:faulttolerance}
We extend the base variant of \name by a fault recovery mechanism.
If one or more $P$s stall (i.e., are faulty) and do not submit their blinded vote in the voting stage despite committing in doing so, the tally cannot be computed directly.
To recover from faulty $P$s, we adapt the solution proposed by Khader et al.~\cite{KhaderSRH12}, and we place the fault recovery phase immediately after the voting phase. 
All remaining honest $P$s are expected to repair their vote by a transaction to $SC$, which contains key materials shared with all faulty $P$s and their NIZK proof of correctness.
$SC$ verifies all key materials with proofs (see  \autoref{fig:ZKPdiffie}), and then they are used to invert out the counter-party keys from a blinded vote of an honest $P$ who sent the vote-repairing transaction to $SC$.

\subsubsection{Faults in Fault Recovery.}
Even if some of the honest (i.e., non-faulty) $P$s would be faulty during the recovery phase (i.e., do not submit vote-repairing transaction), it is still possible to recover from such a state by repeating the next round of the fault recovery, but this time only with key materials related to new faulty $P$s.
Such a partial repair of votes enables us to save costs imposed by the smart contract platform, and thus already submitted key materials need to be neither resubmitted nor verified. 
As a consequence of the fault recovery, it is not necessary to restart the voting protocol, which substantially increases the resistance of our approach to serious failures.
An issue that needs to be addressed in the fault recovery is $\mathcal{A}$ who controls multiple $P$s and let them stall one-by-one in each fault recovery round. 
To disincentivize such a behavior, we require stalling $P$s of each recovery round to lose their deposits, which is split across remaining $P$s as a compensation for the cost of fault recovery -- therefore, the value of the deposit must reflect it.
Moreover, additional (increasing) deposit might be required in each fault-recovery round.

\begin{figure}[t]
	\begin{center}
		\vspace{-0.3cm}
		\fbox{\scriptsize
			\begin{protocolm}{2}	
				
				\participants{\underline{$Participant~P_i$}}{\underline{Smart Contract}}
				\participants{\underline{$A\leftarrow g^{x_i},B
						\leftarrow g^{x_j},x_i$}}{\underline{$A\leftarrow g^{x_i},~B\leftarrow g^{x_j}$}}
				Let~w_i,\in_{r}\mathbb{F}_p\\
				C\leftarrow g^{x_ix_j}\\
				m_1\leftarrow {g}^{w_i}\\
				m_2\leftarrow {B}^{w_i}  \\

				c\leftarrow  \Hzk(A,B,m_1,m_2)\\ 	
				r_i\leftarrow w_i+ cx_i &&\\
				&\sends{C,r_i,m_1,m_2} & c\leftarrow  \Hzk(A,B,m_1,m_2)\\

				&  & g^r\stackrel{?}{=}{m_1}{A^c}\\
				&  & B^r\stackrel{?}{=}{m_2}{C^c}\\

			\end{protocolm}
		}
	\end{center}
	\vspace{-0.35cm}
	\caption{ZKP verifying correspondence of $g^{x_ix_j}$ to public keys $A=g^{x_i}, B=g^{x_j}$. 
	}
	\label{fig:ZKPdiffie}
	\vspace{-0.5cm}
\end{figure}

\myparagraph{\textbf{Example}}
\autoref{fig:linear} depicts two faulty participants $P_3$ and $P_5$.
First, we assume that there is one faulty participant $P_3$ after the voting phase.
Therefore, in the execution of the fault recovery round, $P_3$'s key material must be removed from the blinded votes of all other benign $P$s by inverting out their counter-party key materials $\forall P_i \setminus P_3: g^{x_ix_3}$.
In detail, all benign $P$s are supposed to submit a vote repairing transaction to $SC$ to finish the fault recovery.
However, say $P_5$ (intentionally) stalls during the fault recovery execution.
In this case, after the expiration of timeout for fault recovery round, $P_5$ is also treated as faulty, and all remaining participants $\forall P_i \setminus \{P_3, P_5\}$ must re-execute the fault recovery again, this time only with one key material $g^{x_ix_5}$ related to $P_5$ and its NIZK proof in the arguments of a transaction submitted to $SC$. 
\begin{figure}[h]
	\scriptsize
	\centering		
	\vspace{-1.1cm}
	\begin{equation}
	\boxed{
		\begin{array}{rcl}
		P_1: g^{x_1y_1}\cdot f_1= (g^{-x_1x_2-x_1\underline{x_3}-x_1x_4-x_1\underline{x_5}-x_1x_6})\cdot f_1\\
		P_2:\enspace\; g^{x_2y_2}\cdot f_2=  (g^{x_1x_2-x_2\underline{x_3}-x_2x_4-x_2\underline{x_5}-x_2x_6})\cdot f_2\\
		\triangleright P_3:\enspace\;
		g^{x_3y_3}\cdot f_3=  (g^{x_1x_3+x_2x_3-{x_3}x_4-x_3x_5-x_3x_6})\cdot f_3\\
		P_4: \enspace\;g^{x_4y_4}\cdot f_4=  (g^{x_1x_4+x_2x_4+\underline{x_3}x_4-x_4\underline{x_5}-x_4x_6})\cdot f_4\\
		\triangleright P_5:\enspace\;
		g^{x_5y_5}\cdot f_5=  (g^{x_1x_5+x_2x_5+x_3x_5+x_4{x_5}-x_5x_6})\cdot f_5\\
		P_6: \enspace\;g^{x_6y_6}\cdot f_6= (g^{x_1x_6+x_2x_6+\underline{x_3}x_6+x_4x_6+\underline{x_5}x_6})\cdot f_6\\			
		\end{array}
	}
	\end{equation}
	\vspace{-0.4cm}
	
	\caption{An example of faulty participants $P_3$ and $P_5$. Each honest participant inverts out all components of blinding keys containing underlined private keys of $P_3$ and $P_5$.}
	\label{fig:linear}
			\vspace{-0.5cm}
\end{figure}

The underlined variables in \autoref{fig:linear} show the distribution of ephemeral private keys from both $P_3$ and $P_5$, which are presented in the blinding keys of benign $P$s.
In detail, $SC$ verifies whether the supplied $g^{x_ix_j}$ corresponds to $P_i$'s and $P_j$'s public keys ($g^{x_i}$ and $g^{x_j}$) using the ZKP verification from \autoref{fig:ZKPdiffie}.
Once this phase is completed without new faulty $P$s, the phase of our protocol is shifted to the tally phase, and it continues as before, but without the votes of faulty $P$s.

\section{Implementation \& Evaluation}
\label{sec:implementation}

We selected the Ethereum-based environment for evaluation due to its wide\-spread adoption and open standardized architecture (driven by the Enterprise Ethereum Alliance~\cite{EEA}), which is incorporated by many blockchain projects.
We implemented $SC$ components in Solidity, while $VA$ and $P$ components were implemented in Javascript as testing clients of the \textit{truffle} project.
Executing smart contracts over blockchain, i.e., performing computations and storing data, has its costs. 
In Ethereum Virtual Machine (EVM), these costs are expressed by the level of execution complexity of particular instructions, referred to as \textit{gas}.
In this section, we analyze the costs imposed by our approach, perform a few optimizations, and compare the costs with OVN~\cite{McCorrySH17}.
In the context of this work, we assume 10M as the block gas limit.
With block gas limit assumed for a single transaction, our implementation supports up to 135 participants (see \autoref{fig:mpc-comparison}), up to 7 vote choices (see \autoref{fig:submit-vote}), and up to 9 simultaneously stalling faulty participants (see \autoref{fig:repair-vote}).\footnote{This maximum corresponds to a single recovery round but the total number of faulty participants can be unlimited since the fault recovery round can be repeated.}

We made two different implementations, the first one  is based on DLP for integers modulo $p$ (denoted as integer arithmetic (IA)), and the second one is based on the elliptic curve DLP (denoted as ECC).
We refer the reader to \autoref{sec:appendixecc} for the ECC notation of our work, which is analogous to IA notation from \autoref{sec:protocol}. 
In the ECC, we used a standardized \textit{Secp256k1} curve from existing libraries~\cite{witnet-ecc-solidity}, \cite{McCorrySH17}.
In the case of IA, we used a dedicated library~\cite{bignumber-solidity} for operations with big numbers since EVM natively supports only 256-bit long words, which does not provide sufficient security level with respect to the DLP for integers modulo $p$.\footnote{Since this DLP was already computed for 795-bit long safe prime in 2019~\cite{integer-dlp-2019}, only values higher than 795-bit are considered secure enough.}
We consider 1024 bits the minimal secure (library-supported) length of numbers in IA.
As we will show below, IA implementation even with minimal 1024 bits is overly expensive, and thus in many cases does not fit the block gas limit by a single transaction. 
Therefore, in our experiments, we put emphasis on ECC implementation that is more efficient and preferred option.
The source code of our implementation is available at \url{https://www.dropbox.com/sh/f2gmjnhov70ndjk/AAAqe6bdMKQFgAJ9dBqvTOHda?dl=0}.

\subsection{Cost Optimizations}
\label{ssec:optimizatations}

Since ECC operations in ZKP verifications (see \autoref{fig:ECCZKPMultiCandidate} and \autoref{fig:ZKPECC} of \autoref{sec:appendixecc}) and computation  of MPC keys (see \autoref{eqn:eccmpc} of \autoref{sec:appendixecc}) impose a high execution cost in the setting of decentralized blockchains, we have made several cost optimizations.

\subsubsection{\textbf{(1) Caching in MPC Key Computation.}}
If implemented nai\-vely, the computation of all MPC keys in $SC$ (see \autoref{eqn:eccmpc} of \autoref{sec:appendixecc}) would contain a high number of overlapped additions, 
\begin{wrapfigure}{t}{0.65\textwidth}
	\centering
	\includegraphics[width=\linewidth]{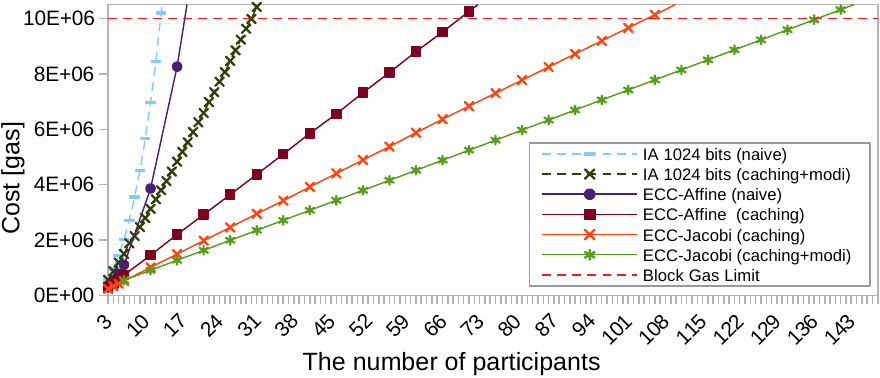}
	\caption{A computation of MPC keys by the authority.} 
	\label{fig:mpc-comparison}
	\vspace*{-0.4cm}
\end{wrapfigure}
and hence the price would be excessively high (see series \textit{``ECC-Affine (naive)''} in \autoref{fig:mpc-comparison}).\footnote{The same phenomenon occurs in IA (see \autoref{eqn:eq1}) but with overlapped multiplications (see series \textit{``IA 1024 bits (naive)''}).}
Therefore, in the code of $SC$ related to the computation of all MPC keys, we accumulate and reuse the value of the left side of \autoref{eqn:eccmpc} during iteration through all participants. 
Similarly, the sum at the right side can be computed when processing the first participant, and then in each follow-up iteration, it can be subtracted by one item. 
However, we found out that subtraction imposes non-negligible costs since it contains one affine transformation (which we later optimize). 
In the result, we found pre-computation of all intermediary right items in the expression during the processing of the first participant as the most optimal. 
The resulting savings are depicted as \textit{``ECC-Affine (caching)''} series in \autoref{fig:mpc-comparison}.
We applied the same optimization for IA; however, even after adding a further optimization (i.e., pre-computation of modular inverses; see \autoref{ssec:optimizatations}.4), the costs were still prohibitively high (see \textit{``IA 1024 bits (caching+modi)''} series in \autoref{fig:mpc-comparison}) and the maximum number of participants fitting the block gas limit was equal only to 29.

\begin{figure}[b]
		\vspace{-1.2cm}
	\centering	
	\subfloat[Vote submission by $P_i$.\label{fig:submit-vote}]{	
		\includegraphics[width=0.48\columnwidth]{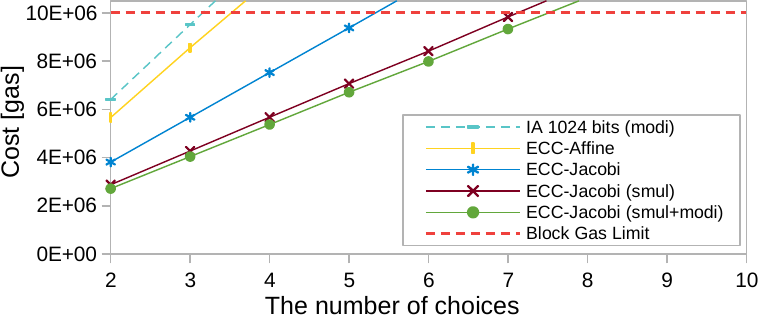}
		\vspace{0.3cm}
	}
	\subfloat[Vote repair by $P_i$.\label{fig:repair-vote}]{	
		\includegraphics[width=0.48\columnwidth]{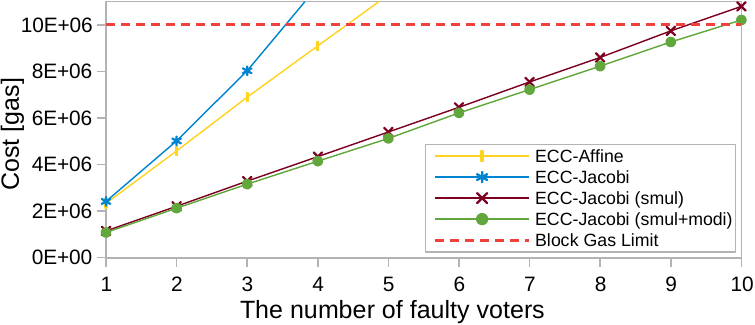}	
		\vspace{0.3cm}
	}		
	\caption{Vote submission and vote repair (i.e., fault recovery) with various optimizations.}
	\label{fig:submit-and-repair-vote}
			\vspace{-0.6cm}
	
\end{figure}

\subsubsection{\textbf{(2) Affine vs. Jacobi Coordinates.}}
In the ECC libraries employed~\cite{witnet-ecc-solidity} \cite{McCorrySH17}, by default, all operations are performed with points in Jacobi coordinates and then explicitly transformed to affine coordinates.
However, such a transformation involves one modular inversion and four modular multiplications over 256-bit long integers, which is costly.
Therefore, we maximized the utilization of internal functions from the Witnet library~\cite{witnet-ecc-solidity}, which do not perform affine transformation after operation execution but keep the result in Jacobi coordinates.
This is possible only until the moment when two points are compared since a comparison of points in Jacobi coordinates can be made only after the points are transformed to affine coordinates. 
Hence, a few calls of the affine transformation are inevitable -- e.g., in the case of ZKP verification during the vote submission (see the $SC$ side in \autoref{fig:ECCZKPMultiCandidate} of \autoref{sec:appendixecc}), there is one affine transformation required in the first check and two affine transformations in the second check.
This optimization is depicted in \autoref{fig:mpc-comparison} (series \textit{``ECC-Jacobi (caching)''}) and \autoref{fig:submit-vote} (series \textit{``ECC-Jacobi''}). 
In the case of computation  of MPC keys, this optimization brought improvement of costs by $23\%$ in contrast to the version with affine coordinates and caching enabled.
Due to this optimization, up to $111$ participants can be processed in a single transaction not exceeding the block gas limit.
In the case of vote submission, this optimization brought improvement of costs by $33\%$ in contrast to the version with affine coordinates.

\subsubsection{\textbf{(3) Multiplication with Scalar Decomposition.}}
The most expensive operation on an elliptic curve is a scalar multiplication; based on our experiments, it is often 5x-10x more expensive than the point addition (in terms of gas consumption) since multiplication itself involves several point additions (and/or point doubling).
The literature proposes several ways of optimizing the scalar multiplication, where one of the most significant ways is w-NAF (Non-Adjacent Form) scalar decomposition followed by two simultaneous multiplications with scalars of halved sizes~\cite{hankerson2005guide}.
This approach was also adopted in the Witnet library~\cite{witnet-ecc-solidity} that we base on.
The library boosts the performance (and decreases costs) by computing the sum of two simultaneous multiplications $kP + lQ$, where $k = (k_1 + k_2 \lambda)$, $l = (l_1 + l_2 \lambda)$, and $\lambda$ is a specific constant to the endomorphism of the elliptic curve.
To use this approach, a scalar decomposition of $k$ and $l$ needs to be computed beforehand.
Nevertheless, such a scalar decomposition can be computed off-chain (and verified on-chain), while only a simultaneous multiplication is computed on-chain.
However, to leverage the full potential of the doubled simultaneous multiplication, one must have the expression $kP + lQ$, which is often not the case.
In our case, such a form fits the second check of vote's ZKP verification at $SC$ (see the right side of \autoref{fig:ECCZKPMultiCandidate} in \autoref{sec:appendixecc}) since the expression of the second check can be rewritten to
$
r_1 H + d_i F_l \stackrel{?}{=} B_l + d_l V_i,
$
which enables us to apply a simultaneous multiplication on the left side.
The first check of the expression in \autoref{fig:ECCZKPMultiCandidate} can be also modified to leverage simultaneous multiplication as follows:
$
r_l G - d_l X  \stackrel{?}{=} A_l.
$
However, in this case, we have to invert both decomposed scalars of $d_l$ to satisfy $d_l = d_l^1 + d_l^2 ~\lambda ~mod~ n$.
Alike the vote submission, this optimization can be applied in vote repair (see \autoref{fig:ZKPECC}), where both checks can be rewritten to their equivalent form containing sums of multiplications: $rG -hA \stackrel{?}{=} m_1$ and $rB -cC  \stackrel{?}{=} m_2$.
We depict the performance improvement brought by this optimization as series \textit{``ECC-Jacobi (smul)''} in \autoref{fig:submit-and-repair-vote}.

\subsubsection{\textbf{(4) Pre-Computation of Modular Inversions.}}\label{sec:optim-modular-inv}
Each affine transformation in the vote submission contains one operation of modular inversion -- assuming previous optimizations, ZKP verification of one item in 1-out-of-$k$ ZKP (see \autoref{fig:ECCZKPMultiCandidate} of \autoref{sec:appendixecc}) requires three affine transformations (e.g., for $k=5$, it is $15$). 
Similarly, the ZKP verification of correctness in the repair vote (see \autoref{fig:ZKPECC} of \autoref{sec:appendixecc}) requires two affine transformations per each faulty participant submitted.
The modular inversion operation runs the extended Euclidean algorithm, which imposes non-negligible costs.
However, all modular inversions can be pre-computed off-chain, while only their verification can be made on-chain (i.e., modular multiplication), which imposes much lower costs.
We depict the impact of this optimization as \textit{``ECC-Jacobi (smul+modi)''} series in \autoref{fig:submit-and-repair-vote} and ``...modi...'' series in \autoref{fig:mpc-comparison}.
In the result, it has brought $5\%$ savings of costs in contrast to the version with the simultaneous multiplication.

\renewcommand{\arraystretch}{0.9}
\setlength{\tabcolsep}{0.1cm}
\begin{table}[t]
	\centering	
	\scriptsize
	\vspace{-0.7cm}
	
	\subfloat[1 core]{	
		\begin{tabular}{r l l l l}		
			\toprule
			\textbf{Voters} & \multicolumn{4}{c}{\textbf{Choices}}  \\ [0.5ex]
			
			(n) & ~$k=2$ & ~$k=4$ & ~$k=6$ & ~$k=8$  \\ [0.5ex]
			\midrule
			\textbf{20} & $0.01s$ & $0.01s$ &  $0.07s$ & $0.07s$\\						
			\textbf{30} & $0.01s$ & $0.01s$ & $0.53s$ & $13.3s$\\								
			\textbf{40} & $0.01s$ & $0.04s$ & $02.6s$ & $112s$\\							
			\textbf{50} &$0.01s$ & $0.08s$ & $10.0s$ & $606s$\\
			
			\textbf{60} &$0.01s$ & $0.16s$ & $28.2s$ & $2424s$\\
			\textbf{70} &$0.01s$ & $0.48s$ & $69.6s$ & $\sim2.1h$\\
			\textbf{80} &$0.01s$ & $0.82s$ & $160s$ & $\sim5.8h$\\
			\textbf{90} &$0.01s$ & $1.08s$ & $320s$ & $\sim14.2h$\\

			\textbf{100}& $0.01s$ & $1.2s$ & $722s$ & $\sim33h$\\
			
			\bottomrule
		\end{tabular}
	}
	\hspace{0.2cm}
	\subfloat[8 cores]{
		
		\begin{tabular}{r l l}		
			\toprule
			\textbf{Voters} & \multicolumn{2}{c}{\textbf{Choices}}  \\ [0.5ex]
			
			(n)  & ~$k=6$ & ~$k=8$  \\ [0.5ex]
			\midrule
			\textbf{20}  &  $0.01s$ & $0.01s$\\						
			\textbf{30}  & $0.08s$ & $2.0s$\\								
			\textbf{40} & $0.39s$ & $16.8s$\\							
			\textbf{50}  & $1.5s$ & $90.9s$\\					
			\textbf{60}  & $4.44s$ & $267s$\\					
			\textbf{70}  & $11.85s$ & $773s$\\								
			\textbf{80}  & $19.46s$ & $2210s$\\								
			\textbf{90}  & $44.02s$ & $\sim2.7h$\\								
			\textbf{100} &  $108.3s$ & $\sim4.9h$\\			
			\bottomrule
		\end{tabular}
	}
	\caption{Upper time bound for tally computation.}	\label{table:bruteforcetime}
	\vspace{-0.9cm}
	
\end{table}

\subsection{Tally Computation}\label{sec:tally-exps}
\vspace{-0.3cm}
In \autoref{table:bruteforcetime}, we provide time measurements of tally computation through the entire search space on a single core vs. all cores of the Intel Core i7-10510U CPU laptop.\footnote{Note that in some cases we estimated the time since we knew the number of attempts.} 
We see that for $n \le 100$ and $k \le 6$, the tally can be computed even on a commodity PC in a reasonable time.
However, for higher $n$ and $k$, we recommend using a more powerful machine or distributed computation across all $P$s. 
One should realize that our measurements correspond to the upper bound, and if some ranges of tally frequencies are more likely than other ones, they can be processed first -- in this way, the computation time can be significantly reduced.
Moreover, we emphasize that an exhaustive search for tally computation is not specific only to our scheme but to homomorphic-encryption-based schemes providing perfect ballot secrecy and privacy of votes (e.g.,~\cite{Kiayias2002},~\cite{HaoRZ10},~\cite{McCorrySH17}).

\renewcommand{\arraystretch}{1.1}
\setlength{\tabcolsep}{0.15cm}
\begin{table}[b]
	\vspace{-0.6cm}
	\centering
	\scriptsize
	
	\begin{tabular}  {l c l l}
		\toprule
		& \textbf{Gas Paid by} & \textbf{OVN} &\textbf{\name}   \\
		\midrule
		
		\textbf{\specialcell{Deployment of Voting $SC$}}   	& $VA$ &  3.78M  & 4.8M  \\
		
		\textbf{\specialcell{Deployment of\\Cryptographic $SCs$}}  & $VA$ &  \specialcell{2.44M}  & \specialcell{2.15M\\(1.22M+0.93M)}   \\
		
		\textbf{Enroll voters} & $VA$ & \specialcell{2.38M\\(2.15M+0.23M)}  & 1.93M \\
		
		\textbf{\specialcell{Submit Ephemeral PK}} & $P$ &  0.76M & 0.15M  \\
		
		\textbf{Cast Vote} & $P$ &  2.50M & 2.72M \\
		
		\textbf{Tally} & $VA ~(or ~P)$ & 0.75M  & 0.39M  \\				
		
		\midrule
		\textbf{Total Costs for} $\mathbf{P}$ &   & 3.26M  & 2.87M \\
		
		\textbf{Total Costs for} $\mathbf{VA}$ &  &  9.35M  & 9.27M  \\

		\bottomrule
		\vspace{-0.2cm}
	\end{tabular}
	\caption{A normalized cost comparison of \name with OVN for $n=40$ and $k=2$.}\label{tab:costcomparisons}
	\vspace{-0.8cm}
\end{table}

\subsection{Cost Comparison}
\vspace{-0.3cm}
In \autoref{tab:costcomparisons}, we made a cost comparison of \name (using ECC) with OVN~\cite{McCorrySH17}, where we assumed two choices and 40 participants (the same setting as in~\cite{McCorrySH17}).
We see that the total costs are similar but \name improves $P$'s costs by $13.5\%$ and $VA$'s cost by $0.9\%$ even though using more complex setting that allows 1-out-of-k voting.
We also emphasize that the protocol used for vote casting in \name \textbf{contains more operations} than OVN but regardless of it, the costs are close to those of OVN, which is mostly caused by the proposed optimizations.\footnote{To verify 1-out-of-$k$ ZKP in vote casting, \name computes $5 \cdot k$ multiplications and $3 \cdot k$ additions on the elliptic curve -- i.e., 10 multiplications and 6 additions for  $k=2$. In contrast, OVN computes only 8 multiplications and 5 additions for $k=2$.}
Next, we found that OVN computes tally on-chain, which is an expensive option.
In contrast, \name computes tally off-chain and $SC$ performs only verification of its correctness, which enables us to minimize the cost of this operation.
Another gas saving optimization of \name in contrast to OVN (and Hao et al.~\cite{HaoRZ10}) is that we do not require voters to submit ZKP of knowledge of $x_i$ in $g^{x_i}$ during the registration phase to $SC$ since $P_i$ may only lose by providing incorrect ephemeral public key $g^{x_i}$ -- she might lose the chance to vote and her deposit.
Finally, we note that we consider the deployment costs of our $SC$ equal to 4.8M units of gas; however, our $SC$ implementation contains a few auxiliary view-only functions for a pre-computation of modular inverses, with which, the deployment costs would increase to 7.67M due to code size.
Nevertheless, these operations can be safely off-chained and we utilized them on-chain only for simplicity.

\section{Discussion of Extensions}
\label{sec:discussion}
\vspace{-0.3cm}
In this section, we discuss the extensions addressing the scalability and performance limitations of \name. 

\vspace{-0.2cm}
\subsection{Scalability Limitation \& Extension}
\label{ssec:scalability-improvements}
\vspace{-0.2cm}
The limitation of \name (like in OVN) is a lack of scalability, where the block gas limit might be exceeded with a high number of $P$s.
Therefore, we primarily position our solution as boardroom voting; however, we will  show in this section that it can be extended even to larger voting.
Our voting protocol (see \autoref{ssec:basicprotocol}) has a few platform-specific bottlenecks. 
The first bottleneck is in the setup phase, where $VA$ submits the wallet addresses of all $P$s to SC in a single transaction, which might exceed the block gas limit when the number of participants $n > 201$.
The second bottleneck occurs in the pre-voting phase, where $VA$ calls the function of $SC$ to compute all MPC keys; exceeding the block gas limit occurs when the number of participants $n > 135$. 
The next bottleneck occurs in the voting phase, 
where voters submit their blinded votes together with 1-out-of-$k$ ZKP,
exceeding the block gas limit when the number of choices $k > 7$.
The last bottleneck occurs in the fault recovery phase and the block gas limit is exceeded when the number of simultaneously faulty participants $f > 9$.

\begin{figure}
	\centering
	\includegraphics[width=0.7\linewidth]{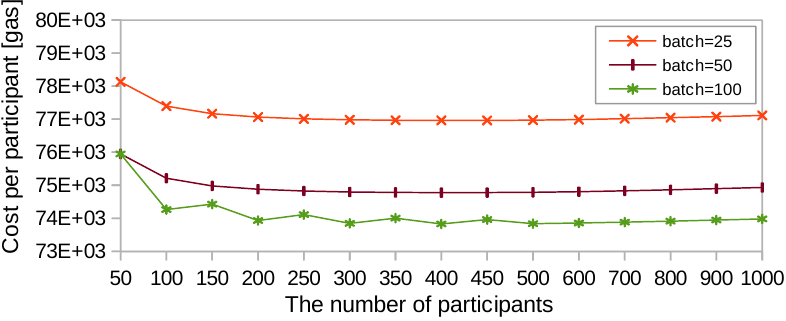}
	\caption{A computation of MPC keys by the authority $VA$ using various batch sizes and the most optimized ECC implementation, demonstrating a scalability extension of \name, which also mitigates a privacy issue of unanimous and majority voting.}
	\label{fig:mpc-batches}
\end{figure}
Nevertheless, transaction batching can be introduced for the elimination of all these bottlenecks.
To realize a batching of pre-voting, voting, and fault-recovery  phases, the additional integrity preservation logic across batches needs to be addressed while the verification of integrity has to be made by $SC$.
For demonstration purposes, we addressed the bottleneck of the pre-voting stage (see \autoref{fig:mpc-batches})  and setup stage, 
which further improves the vote privacy in \name (see \autoref{sec:blockchain-specific-issues}) and causes only a minimal cost increase due to the overhead of integrity preservation (i.e., 1\%).
With this extension, $n$ is limited only by the expenses paid by $VA$ to register $P$s, compute their MPC keys, and its computing power to obtain the tally.\footnote{E.g., for $n = 1000$, $k=2 ~(\text{and} ~k=4)$, it takes $0.15$s (and $\sim4$h) to obtain the tally on a commodity PC with 8 cores, respectively.}
If a certain combination of high $n$ and $k$ would make computation of a tally overly computationally expensive (or the cost of its verification by $SC$), it is further possible to partition $P$s into multiple groups (i.e., voting booths), each managed by an instance of \name, while the total results could be summed across instances.
Scalability evaluation of such an approach and other operations is subject to our our other work which provides more empirical evidence~\cite{stanvcikova2022sbvote}.

\subsection{Cost and Performance Limitations}\label{ssec:cost-perf-limitations}
\vspace{-0.3cm}
Although we thoroughly optimized the costs (and thus performance) of our implementation (see \autoref{ssec:optimizatations}), the expenses imposed by a public permissionless smart contract platform might be still high, especially during peaks of the gas price and/or in the case of a larger voting than boardroom voting (see \autoref{ssec:scalability-improvements}).
Besides, the transactional throughput of such platforms might be too small for such larger voting instances to occur in a specified time window.
Therefore, to further optimize the costs of \name and its performance, it can run on a 
public permissioned Proof-of-Authority platforms, e.g., using Hyperledger projects (such as safety-favored Besu with BFT).
Another option is to use smart contract platforms backed by the trusted computing that off-chains expensive computations (e.g., Ekiden~\cite{cheng2019ekiden}, TeeChain~\cite{lind2019teechain}, Aquareum~\cite{homoliak2020aquareum}), or other partially-decentralized second layer solutions (e.g., Plasma~\cite{plasma-eth}, Polygon Matic~\cite{polygon-matic-investopedia}, Hydra~\cite{chakravarty2020hydra}). 
Even though these solutions might preserve most of the blockchain features harnessed in e-voting, availability and decentralization might be decreased, which is part of the security/performance trade-off.

\section{Security Analysis}
\label{ssec:security}	
\vspace{-0.2cm}
In this section, we first analyze security of \name with regard to the voting properties specified in \autoref{sec:preliminaries}.
Next, we analyze blockchain-specific security \& privacy issues and discuss their mitigations. 
The security of our scheme relies on well-known cryptographic primitives under their standard assumptions.

\subsection{Properties of Voting}
\vspace{-0.3cm}
\par\noindent\textbf{(1) Privacy}
in \name requires at least 3 $P$s, out of which at least 2 are honest (see the proof in \autoref{sec:appendixvoteorivacy}).
Privacy in BBB-voting is achieved by blinding votes using ElGamal encryption~\cite{Elgamal85}, whose security is based on the decisional Diffie-Hellman assumption. Unlike the conventional ElGamal algorithm, a decryption operation is not required to unblind the votes. 
Instead, we rely on the self-tallying property of our voting protocol. 
The ciphertext representing a blinded vote is a tuple ($c_1,c_2)$, where $c_1=g^{xy}.f$ and $c_2=g^y$, where the purpose of $c_2$ is to assist with the decryption.
Decryption involves computing $(c_2)^{-x}\cdot c_1$ to reveal $f$, which unambiguously identifies a vote choice. 
As a result, the blinding operation for participant $P_i$ in \autoref{eqn:eq2} is equivalent to ElGamal encryption involving the computation of $c_1$ but not the decryption component $c_2$.
Furthermore, the blinding keys are ephemeral and used exactly once for encryption (i.e., blinding) of the vote within a single run of voting protocol 
-- i.e., if the protocol is executed correctly, there are no two votes $f_l$ and $f_m$ encrypted with the same ephemeral blinding key of $P_i$, such that 
\begin{equation}
 \frac{(g^{xy}\cdot f_l)} {(g^{xy}\cdot f_m)} = \frac{f_l}{f_m},  
\end{equation}
from which the individual votes could be deduced.  
For the blockchain-specific privacy analysis, see also (2) and (3) of \autoref{sec:blockchain-specific-issues}.

\par\noindent\textbf{(2) Ballot Secrecy.} It is achieved by blinding the vote using ElGamal homomorphic encryption~\cite{cgs97}, and it is not required to possess a private key to decrypt the tally because of the self-tallying property ($g^{\sum_{i}x_i y_i}=1$).
Therefore, given a homomorphic encryption function, it is possible to record a sequence of encrypted votes without being able to read the votes choices. 
However, if all $Ps$ are involved in the recovery of a partial tally consisting of a recorded set of votes, these votes can be unblinded (as allowed by ballot secrecy).
Even a subset of $n-2$ $P$s\footnote{Note that at least 2 $P$s are required to be honest (see \autoref{sec:adversary-model}).}  who have already cast their votes cannot recover a partial tally that reveals their vote choices because of the self-tallying property ($g^{\sum_{i}x_i y_i}=1$) has not been met.

\par\noindent\textbf{(3) Fairness}. 
If implemented naively, the last voting $P$ can privately reveal the full tally by solving \autoref{eqn:eq3} before she casts her vote since all remaining blinded votes are already recorded on the blockchain (a.k.a., the last participant conundrum).
This can be resolved by $VA$ who is required to submit the final dummy vote including the proof of her vote choice, which is later subtracted from the final tally by $SC$.\footnote{Note that If $VA$ were for some reason not to execute this step, the fault recovery would exclude $VA$'s share from MPC keys, and the protocol would continue.}

\par\noindent\textbf{(4a) Universal Verifiability}. 
Any involved party can check whether all recorded votes in the blockchain are correct, and they are correctly included in the final tally~\cite{Kiayias2002}.
Besides, the blinded votes are verified at $SC$, which provides correctness of its execution and public verifiability, relying on the honest majority of the consensus power in the blockchain (see \autoref{sec:blockchain-specific-issues}).

\par\noindent\textbf{(4b) E2E Verifiability}. 
To satisfy E2E verifiability~\cite{EPRINT:PanRoy18}, two properties should be met: 
(I) each $P$ can verify whether her vote was cast-as-intended and recorded-as-cast, 
(II) anyone can verify whether all votes are tallied-as-recorded.
\name meets (I) since each $P$ can locally compute her vote choice (anytime) and compare it against the one recorded in the blockchain.\footnote{Since $P$ can anytime (possibly upon coercion) create a receipt based on her vote choice (that cannot be changed), receipt-freeness is not provided.
}
\name meets (II) since $SC$ executes the code verifying that the submitted tally fits \autoref{eqn:eq3} that embeds all recorded votes in the blockchain.

\par\noindent\textbf{(5) Dispute-Freeness}. 
Since the blockchain acts as a tamper-resistant bulletin board (see \autoref{sec:blockchain-specific-issues}), and moreover it provides correctness of code execution (i.e., on-chain execution of verification checks for votes, tally, and fault recovery shares) and verifiability, the election remains dispute-free under the standard blockchain assumptions about the honest majority and waiting the time to finality.

\par\noindent\textbf{(6) Self-Tallying}. 
\name meet this property since in the tally phase of our protocol (and anytime after), all cast votes are recorded in $SC$; therefore any party can use them to fit \autoref{eqn:eq3}, obtaining the final tally.

\par\noindent\textbf{(7) Robustness (Fault Tolerance)}. 
\name is robust since it enables to remove (even reoccurring) stalling $P$s by its fault recovery mechanism (see \autoref{sec:faulttolerance}).
Removing of stalling $P$s involves $SC$ verifiability of ZKP submitted by $P$s along their counter-party shares corresponding to stalling $P$s.

\par\noindent\textbf{(8) Resistance to Serious Failures}. 
The resistance of \name to serious failures relies on the integrity and append-only features of the blockchain, which (under its assumptions \autoref{sec:blockchain-specific-issues}) does not allow the change of already cast votes.

\subsection{Blockchain-Specific Aspects and Issues}\label{sec:blockchain-specific-issues}
\vspace{-0.3cm}
In the following, we focus on the most important blockchain-specific aspects and issues related to e-voting with \name.
For analysis of other potential issues and their comparison with centralized e-voting, we refer the reader to \autoref{sec:appendix-blockchain-vs-centralized-voting}, where we conclude that 
blockchain e-voting (with \name as the example) can only improve the potential problems of centralized e-voting (such as censorship, DoS, device exploitation) at the cost of more complex governance.

\par\noindent\textbf{(1) Bulletin Board vs. Blockchains}. 
The definition of a bulletin board~\cite{Kiayias2002} assumes its immutability and append-only feature, which can be provided by blockchains that moreover provide correct execution of code.

CAP theorem~\cite{brewer2000towards} enables a distributed system (such as the blockchain) to select either \textbf{\underline{c}}onsistency or  \textbf{\underline{a}}vailability during the time of network \textbf{\underline{p}}artitions.
If the system selects consistency (e.g., Algorand~\cite{gilad2017algorand}, BFT-based blockchains such as~\cite{hyperledger1}), it stalls during network partitions and does not provide liveness (i.e., the blocks are not produced) but provides safety (i.e., all nodes agree on the same blocks when some are produced). 
On the other hand, if the system selects availability (e.g., Bitcoin~\cite{Satoshi2009}, Ethereum~\cite{wood2014}), it does not provide safety but provides liveness, which translates into possibility of creating \textit{accidental forks} and eventually accepting one as valid.
Many public blockchains favor availability over consistency, and thus do not guarantee immediate immutability. 
Furthermore, blockchains might suffer from \textit{malicious forks} that are longer than accidental forks and are expensive for the attacker. 
Usually, their goal is to execute double-spending or selfish mining~\cite{eyal2018majority}, violating the assumptions of the consensus protocol employed -- more than 51\% / 66\% of honest nodes presented in PoW / BFT-based protocols.
To prevent accidental forks and mitigate malicious forks in liveness-favoring blockchains, it is recommended to wait for a certain number of blocks (a.k.a., block confirmations / a time to finality). 
Another option to cope with forks is to utilize blockchains favoring safety over liveness (such as~\cite{gilad2017algorand,hyperledger1}). 

Considering \name, we argue that these forks are not critical for the proposed protocol since any transaction can be resubmitted if it is not included in the blockchain after a fork.
Waiting for the time to finality (with a potential resubmission) can be done as a background task of the client software at $P$s' devices, so $P$s do not have to wait.
Finally, we emphasize that the time to finality is negligible in contrast to timeouts of the protocol phases; therefore, there is enough time to make an automatic resubmission if needed.

\vspace{-0.1cm}
\par\noindent\textbf{(2) Privacy of Votes}. 
In \name, the privacy of vote choices can be ``violated'' only in the case of unanimous voting by all $P$s, assuming $\mathcal{A}$ who can link the identities of $P$s (approximated by their IP addresses) to their blockchain addresses by  passive monitoring of network traffic.
However, this is the acceptable property in the class of voting protocols that provide the full tally of votes at the output, such as \name and other protocols (e.g.,~\cite{McCorrySH17,Kiayias2002,HaoRZ10,killerprovotum,yu2018platform}).
Moreover, $\mathcal{A}$ can do deductions about the probability of selecting a particular vote choice by $P$s.
For example, in the case that the majority $m$ of all participants $n$ voted for a winning vote choice, then $\mathcal{A}$ passively monitoring the network traffic can link the blockchain addresses of $Ps$ to their identities (i.e., IP addresses), 
and thus $\mathcal{A}$ can infer that each $P$ from the group of all $P$s cast her vote to the winning choice with the probability equal to $\frac{m}{n} > 0.5$.
However, it does not violate the privacy of votes and such an inferring is not possible solely from the data publicly stored at the blockchain since it stores only blinded votes and blockchain addresses of $P$s, not the identities of $P$s.
To mitigate these issues, $P$s can use anonymization networks or VPN services for sending transactions to the blockchain. 
Moreover, neither $\mathcal{A}$ nor $VA$ can provide the public with the indisputable proof that links $P$'s identity to her blockchain address since $P$s never provide $VA$ with such a proof (i.e., all communication of the identity verification is carried out without requiring $P$ to sign such a proof).

\vspace{-0.1cm}
\par\noindent\textbf{(3) Privacy of Votes in Larger than Boardroom Voting}. 
The privacy issue of unanimous and majority voting (assuming $\mathcal{A}$ with network monitoring capability) are less likely to occur in the larger voting than boardroom voting since the voting group of $P$s is larger and potentially more divergent.
We showed that \name can be extended to such a large voting by integrity-preserving batching in \autoref{ssec:scalability-improvements}. 
We experimented with batching up to 1000 $P$s, which is a magnitude greater voting than the boardroom voting.
We depict the gas expenses paid by $VA$ (per $P$) in \autoref{fig:mpc-batches}, where we distinguish various batch sizes.
Results show that the bigger the batch size, the lower the price per~$P$.

\vspace{-0.3cm}
\section{Related Work}
\label{sec:relatedwork}
\vspace{-0.3cm}
In this section, we briefly survey existing paradigms in e-voting and describe a few blockchain-based e-voting approaches. 
In particular, we focus on remote voting approaches (with sufficient specification), which we compare in \autoref{tab:comparison}.

\myparagraph{\textbf{E-Voting Paradigms}}
Utilization of mix-nets that shuffle the votes to break the map between $P$s and their votes was proposed by Chaum~\cite{Chaum1981}.
Benaloh and Fischer~\cite{Cohen1985} were among the first who showed a paradigm shift from anonymizing $P$s to providing privacy of the vote.
Cramer et al.~\cite{cgs97} present a model where all votes are sent to a single combiner, utilizing homomorphic properties of the ElGamal cryptosystem~\cite{Elgamal85}. 
Using bulletin board and zero-knowledge proofs allow their protocol to be universally verifiable.
The work of Kiayias and Yung~\cite{Kiayias2002} converts this scheme into a self-tallying combiner supporting 1-out-of-2 choices; further, the authors outline an extension of their base protocol to support 1-out-of-k choices.
Hao et al.~\cite{HaoRZ10} improve upon the self-tallying protocol by proposing a simple general-purpose two-round voting protocol for 1-out-of-2 choices with low bandwidth requirements and computational costs.
Khader et al.~\cite{KhaderSRH12} take it a step further by adding fairness and  robustness properties.
Groth~\cite{Groth2004} introduces an anonymous broadcast channel with perfect message secrecy leveraged in his voting protocol that is simpler and more efficient than~\cite{Kiayias2002}. 
However, it requires sequential voting, where each voter has to download a fresh state of the bulletin board before voting.
Zagorski et al.~\cite{zagorski2013remotegrity} propose Remotegrity that is based on Scanintegrity~\cite{chaum2008scantegrity} ballots mailed to voters, allowing them remotely vote and verify that their ballots were correctly posted to the bulletin board and at the same time providing protection against malware in clients.
Another direction (e.g.,~\cite{kusters2020ordinos,huber2022kryvos}) focuses on the tally-hiding~\cite{benaloh1986improving,hevia2002electronic} property that is suitable for some use cases requiring to reveal only the best $m$ candidates.

\vspace{-0.2cm}
\myparagraph{\textbf{Location}}
Voting systems can be classified by the physical location where the vote is cast.
Some schemes allow $P$s to submit a vote from their devices (i.e., \textit{remote voting}), e.g.,~\cite{KiayiasAdder06},~\cite{Adida08},~\cite{zagorski2013remotegrity}, and blockchain-based~\cite{followMyVote-blockchain-simple-voting},~\cite{tivi-blockchain-simple-voting}. 
Others systems require voting to be carried out at a designated site (a.k.a., \textit{supervised voting}), e.g.,~\cite{chaum2008scantegrity},~\cite{Rui2012},~\cite{Sandler08},~\cite{Bell2013},~\cite{ShahandashtiH16}, and blockchain-based~\cite{soud2020trustvote},~\cite{hjalmarsson2018blockchain},~\cite{agora2018}.

\vspace{-0.3cm}
\subsection{Blockchain-Based E-Voting}
\vspace{-0.3cm}
We extend the categorization of (remote) blockchain-based voting systems~\cite{yu2018platform} in the following, while we focus on smart contract-based systems.

\vspace{-0.3cm}
\myparagraph{\textbf{(a) Voting Systems Using Smart Contracts}}
McCorry et al.~\cite{McCorrySH17} proposed OVN, a self-tallying voting protocol (basing on~\cite{HaoRZ10}) that provides vote privacy and supports two vote choices.
OVN is implemented as Ethereum $SC$ and is suitable for boardroom voting. 
It does not provide robustness and expensive tally computation is made by $SC$. 
In contrast, \name only verifies the tally in $SC$, while it is computed off-chain.
Similar approach basing on~\cite{HaoRZ10} was proposed by Li et al.~\cite{Li2020}, who further provided robustness from~\cite{KhaderSRH12}.
Seifelnasr et al.~\cite{DBLP:conf/fc/SeifelnasrGY20} aimed to increase the scalability of OVN by off-chaining tally computation and registration at $VA$ in a verifiable way. 
Due to the higher costs imposed by storing data on $SC$, they compute the Merkle tree of voter identities and store only its root hash at $SC$. 
Their approach requires only a single honest $P$ to maintain the protocol's security by enabling her to dispute the incorrect tally submitted to $SC$.
The scalability technique proposed in this paper is orthogonal to us, and it can be combined with our techniques (see \autoref{ssec:scalability-improvements}) to achieve higher savings of the on-chain costs.
Yu et al.~\cite{yu2018platform} employ ring signature to ensure that the ballot is from one of the valid choices, and they achieve scalability by linkable ring signature key accumulation. 
Their approach provides receipt-freeness 
under the assumption of trusted $VA$.
However, we argue that this property is useless when remaining coercion-prone (typical to remote voting).
Moreover, due to receipt-freeness, this approach does not provide E2E verifiability (i.e., cast-as-intended and recorded-as-cast). 
Killer et al.~\cite{killerprovotum} present an E2E verifiable remote voting scheme with two vote choices. 
The authors employ threshold cryptography for achieving robustness using a scheme similar to Shamir secret sharing. 
However, it supports only integers up to a 256 bits (i.e., a size of the EVM word), which is far below a minimal secure length.
Matile et al.~\cite{ICBC:MRSS19} proposed a voting system providing cast-as-intended (but neither E2E nor universal) verifiability. 
Their system uses ElGamal encryption based on DLP with integers modulo $p$. 
Since existing blockchains support natively only up to 256-bit security for this DLP, the authors create the custom blockchain with sufficient security. 
Dagher et al.~\cite{icissp:DMMM18} proposed BroncoVote, a voting system that preserves vote privacy by homomorphic encryption (i.e., Paillier cryptosystem) -- the authors do not provide full deployment on the blockchain but off-chain all cryptographic operations to a trusted server (without verifiability), which introduces a vulnerability. 
Kostal et al.~\cite{kovst2019blockchain} propose voting system, in which $VA$ serves as a trusted key generator that distributes private keys for homomorphic encryption to voters, which enables them to resolve robustness issues at the cost of putting a strong trust into $VA$.
This, however, enables $VA$ to sabotage voting or decrypt the tally before voting finishes.

\vspace{-0.3cm}
\myparagraph{\textbf{(b) Voting Systems Using Cryptocurrency}}
Zhao and Chan~\cite{zhao2015vote} propose a privacy-preserving voting system with 1-out-2 choices based on Bitcoin, which uses a lottery-based approach with an off-chain distribution of voters' secret random numbers with their ZKPs.
The authors use deposits incentivizing $P$s to comply with the protocol; however, a malicious $P$ can sabotage the voting by refusing to vote or vote in a different order than required. 
Tarasov and Tewari~\cite{tarasov2017internet} proposed a conceptual voting system based on Zcash. 
The voter's anonymity (and thus privacy of vote) is ensured by the z-address that preserve unlinkability. 
The correctness of the voting is guaranteed by the trusted $VA$ and the candidates. 
If $VA$ is compromised, double-voting or tracing the source of the ballot (violating privacy) is possible.
Liu and Wang~\cite{liu2017voting} propose a conceptual voting approach that is based on blind signatures with 2 vote choices. 
They utilize blockchain only for (auditable) sending of the messages among parties.
However, despite using blind signatures, $P$s send their vote to blockchain in plain-text, thus the authors recommend anonymization services.

\vspace{-0.3cm}
\myparagraph{\textbf{(c) Commercial Voting Systems with Ballot Box}}
FollowMyVote~\cite{followMyVote-blockchain-simple-voting} and Tivi~\cite{tivi-blockchain-simple-voting} use blockchain for recording of plain text votes into a \textit{ballot box}, where privacy of votes relies only on the assumption that a public key of $P$ cannot be linked to her identity, which is not resistant to $\mathcal{A}$.
NetVote~\cite{netvote2018} is the solution addressing privacy by trusting $VA$ to reveal the encryption key of ballots after the election ends. 
However, $VA$ can refuse to reveal the key or reveal it before the voting ends.
In NetVote, ballot boxes correspond to electoral districts, each of them having its smart contract.
Each such smart contract includes a list of votes, also represented by
smart contracts, which, however, is an expensive solution.

\renewcommand{\arraystretch}{1.1}
\setlength{\tabcolsep}{7.0pt}
\begin{table*}[t]
	\vspace{-0.4cm}
	\centering
	\scriptsize
	
	\begin{tabular}[t]  {l c c c c c c c c c c}
		\toprule
		\textbf{\specialcell{Approach\\}} & \trot{\textbf{Privacy of Votes}} & \trot{\textbf{\specialcell{Perfect Ballot Secrecy}}} & \trot{\textbf{Fairness}}  & \trot{\textbf{Self-Tallying}} & \trot{\textbf{Robustness}} &  \trot{\textbf{\specialcell{ Uses Blockchain}}} & \trot{\textbf{Uni. Verifiability}} &  \trot{\textbf{E2E Verifiability}}  & \trot{\textbf{Open Source}} & \trot{\textbf{Choices}}\\
		\toprule
		
		Hao et al.~\cite{HaoRZ10} & \hfil\cmark & \hfil\cmark & \hfil\xmark & \hfil\cmark & \hfil\xmark & \hfil\xmark  & \hfil\cmark & \hfil\cmark  & \hfil\xmark &  \hfil $2$\\
		
		Khader et.~\cite{KhaderSRH12} & \hfil\cmark & \hfil\cmark & \hfil\cmark & \hfil\cmark & \hfil\cmark  & \hfil\xmark  & \hfil\cmark & \hfil\cmark & \hfil\xmark & \hfil $2$\\
		
		Kiayias and Yung~\cite{Kiayias2002} & \hfil\cmark & \hfil\cmark & \hfil\cmark  & \hfil\cmark & \hfil\cmark & \hfil\xmark  & \hfil\cmark & \hfil\cmark  & \hfil\xmark & \hfil $2 / k$\\
		
		McCorry et.~\cite{McCorrySH17} (OVN) & \hfil\cmark & \hfil\cmark & \hfil\cmark  & \hfil\cmark & \hfil\xmark &  \hfil\cmark & \hfil\cmark & \hfil\cmark  & \hfil\cmark & \hfil $2$\\
		
		Seifelnasr et al.~\cite{DBLP:conf/fc/SeifelnasrGY20} (sOVN) & \hfil\cmark & \hfil\cmark & \hfil\cmark & \hfil\cmark & \hfil\xmark & \hfil\cmark & \hfil\cmark & \hfil\cmark & \hfil\cmark & \hfil $2$  \\

		Li et al.~\cite{Li2020} & \hfil\cmark & \hfil\cmark & \hfil\cmark & \hfil\cmark & \hfil\cmark &  \hfil \cmark  & \hfil\cmark & \hfil\cmark & \hfil\xmark & \hfil $2$\\

		Baudron et al.~\cite{Baudron2001} & \hfil\cmark & \hfil\xmark & \hfil\xmark & \hfil\xmark & \hfil\cmark &  \hfil\xmark & \hfil\cmark & \xmark & \hfil\xmark & \hfil $k$\\ 
		
		Groth~\cite{Groth2004} & \hfil\cmark & \hfil\cmark & \hfil\cmark & \hfil\cmark & \hfil\xmark & \hfil\xmark  & \hfil\cmark & \cmark & \hfil\xmark & \hfil $k$\\
		
		Adida~\cite{Adida08} (Helios) & \hfil\cmark$^*$  & \hfil\xmark & \hfil\xmark & \hfil\xmark & \hfil\cmark & \hfil\xmark & \hfil\cmark & \cmark & \hfil\cmark & \hfil $k$  \\
		
		Matile et al.~\cite{ICBC:MRSS19} (CaIV) & \hfil\cmark & \hfil\xmark & \hfil\xmark & \hfil\xmark & \hfil\cmark & \hfil\cmark & \hfil\xmark & \hfil\xmark & \hfil\cmark &  \hfil $k$\\
		
		Killer~\cite{killerprovotum} (Provotum) & \hfil \cmark & \xmark & \cmark & \xmark & \cmark & \cmark & \cmark & \cmark & \hfil \cmark & $2$ \\

		Dagher et al.~\cite{icissp:DMMM18} (BroncoVote)			 & \hfil \cmark & \cmark$^*$ & \cmark$^*$ & \xmark & \xmark & \cmark & \xmark & \xmark & \xmark & \hfil $k$  \\

		Kostal et al.~\cite{kovst2019blockchain} & \cmark$^*$  &  \cmark$^*$ & \xmark & \cmark & \cmark$^*$ & \cmark & \cmark  & \cmark & \xmark & $k$  \\

		Zagorski et al.~\cite{zagorski2013remotegrity} (Remotegrity)		 & \cmark$^*$ & \xmark & \xmark & \xmark & \cmark & \xmark & \xmark & \cmark & \cmark & $k$  \\

		Yu et al.~\cite{yu2018platform}	& \hfil\cmark & \hfil\cmark$^*$ & \hfil\cmark$^*$ & \hfil\cmark & \hfil\cmark$^*$ & \hfil\cmark & \hfil\cmark & \hfil\xmark  &  \hfil\xmark & \hfil $k$  \\

		\textbf{BBB-Voting} & \hfil\cmark & \hfil\cmark & \hfil\cmark & \hfil\cmark & \hfil\cmark &  \hfil\cmark & \hfil\cmark & \hfil\cmark & \hfil\cmark & \hfil $k$\\
		
		\bottomrule
	\end{tabular}
	\vspace{5pt}
	\caption{A comparison of various remote voting protocols. $^*$Assuming a trusted $VA$.}
	\label{tab:comparison}
	\vspace{-0.8cm}
\end{table*}

\section{Conclusion}
\label{sec:conclusion}
\vspace{-0.3cm}
In this paper, we proposed \name, a 1-out-of-$k$ blockchain-based boardroom voting solution that supports fault tolerance. 
We made two variants of full implementation on the Ethereum -- one based on elliptic curve DLP and the other one based on DLP for integer modulo $p$.
We showed that only the elliptic curve variant is feasible in the real settings of public blockchains.
We performed several cost optimizations and discussed further improvements concerning costs and scalability, where we made a proof-of-concept implementation with up to 1000 voters.
Finally, we compared our solution with OVN and results indicate that \name reduces the costs for voters and the authority by $13.5\%$ and $0.9\%$, respectively; while additionally offering robustness.

\bibliographystyle{splncs04}
\bibliography{ref}

\appendix

\section{Blockchains and Smart Contracts}
\label{sec:appendix-blockchains}	
The blockchain is a data structure representing an append-only distributed ledger
that consists of entries (a.k.a., \textit{transactions}) aggregated within ordered blocks.
The order of the blocks is agreed upon by mutually untrusting participants running a Byzantine fault tolerant consensus protocol -- these participants are different than participants of the voting protocol, and therefore we will refer to them as consensus nodes (a.k.a., miners / validators).
The blockchain is resistant against modifications by design since blocks are linked using a cryptographic hash function, and blocks are considered irreversible after elapsing a time to finality (see \autoref{sec:discussion}).

Each block may contain orders transferring crypto-tokens, application codes written in a platform-supported language, and the execution orders of such applications. 
These application codes are referred to as \textit{smart contracts} and can encode arbitrary processing logic written in the supported language (e.g., Solidity, LLL, Flint,  Scilla, etc).
Interactions between clients and the blockchain are made by messages called \textit{transactions}, which can contain either the orders transferring crypto-tokens or calls of smart contract functions.
All transactions sent to the blockchain are executed and validated by all consensus nodes.
In this way, a smart contract platforms enable decentralized trusted execution of the deployed code, relying on the honest majority of consensus nodes.
We discuss implications of this assumption and other immutability-related properties of blockchains on blockchain-based voting in \autoref{sec:discussion}. 

\section{Elliptic Curve Analogy}
\label{sec:appendixecc}

In this section, we present the analogy of the main steps in the basic protocol implemented on the elliptic curve \textit{Secp256k1}.
The recommended parameters by Brown~\cite{Brown2010} are considered.
The same curve parameters are also used in Bitcoin~\cite{nakamoto2008bitcoin} and Ethereum~\cite{wood2014} for digital signing~\cite{eccparam2019}.
Let $G$ be a base point on the elliptic curve in $\mathbb{F}_p$ of order $nn$ and co-factor $hh=1$.
Let the point at infinity \textit{O} be denoted as 0.

\noindent\textbf{Ephemeral Key Generation}. This step involves generating an ephemeral keypair for each participant.
The ephemeral private key for $P_i$ is a randomly chosen  $x_i\in_{R} \mathbb{Z}_{nn}$ and the ephemeral public key is ${x_i\cdot G}$.

\noindent \textbf{MPC Key Computation.} MPC is used to synchronize the keys between all participants and achieve the \textit{self-tallying} property of the voting.
Each participant is then able to compute
\begin{equation}
\label{eqn:eccmpc}
H_i={y_i}\cdot G=\sum\limits_{j=1}^{i-1} {x_j}\cdot G - \sum\limits_{j=i+1}^{n} {x_j}\cdot G
\end{equation}
The ephemeral blinding key of $P_i$ is computed as $x_i\cdot H_i$, where $H_i$ is found in \autoref{eqn:eccmpc}.
By the self-tallying property, $\sum\limits_{i=1}^{n} x_i\cdot H_i=0$.

\noindent\textbf{Vote Packing, Blinding, and Verification}. This step ensures the recoverability of the tally, vote privacy, and its well-formedness.
Let $\{F_1,F_2,...,F_k\}$ be $k$ independent generator points on the curve that generates the same subgroup as $G$.
The blinded vote options for a participant $P_i$ are
\begin{equation}
\label{eqn:ECCBlindedVoteChoices}
V_i=\begin{cases}
{x_iy_iG}+ F_1 & \text{if $P_i$ votes for candidate 1}\\
{x_iy_iG}+ F_2 & \text{if $P_i$ votes for candidate 2}\\
...\\
{x_iy_iG}+ F_k & \text{if $P_i$ votes for candidate \textit{k}}
\end{cases}
\end{equation}

\begin{figure}[t]
	\begin{center}
		\fbox{\scriptsize
			\begin{protocolm}{2}						
				\participants{\underline{Participant $P_i$}}{\underline{Smart Contract}}
				\participants{($~H\leftarrow {y_i}G,~ v_i$)}{($~H\leftarrow {y_i}G$)}
				\hline	
				& &\\		
				P_i~selects~v_i\in\{1,...,k\}, & & \\	 				
				\text{Use~choice~generators}\\
				F_l\in\{F_1,...,F_k\},\\			
				Publish~ X\leftarrow {x_i}G && 	\\
				and~V_i\leftarrow {x_i}{H}+{F_l} & & \\

				Let~w\in_{R} \mathbb{Z}_{nn}\\

				\forall l\in \{1,..,k\}\setminus {v_i}:\\
				
				\quad 1.~r_l,d_l\in_{R} \mathbb{Z}_{nn} \\
				\quad 2.~A_l\leftarrow {-d_l}X +{r_l}G\\ 				
				\quad 3.~B_l\leftarrow {r_l}H+d_lF_l \\
				\quad\quad\quad\quad-d_lV_i &&\\

				for\enspace {v_i}:\\
				
				\quad 1.~A_{v_i}\leftarrow {w}G\\
				\quad 2.~B_{v_i}\leftarrow {w}H \\
				\hdashline
				c\gets\Hzk(\{ \{A_l\},\{B_l\} \}_{l})\\
				\hdashline
				
				for\enspace {v_i}:\\
				
				\quad 1.~d_{v_i}\leftarrow \sum_{l\neq{v_i}}d_l\\
				\quad 2.~d_{v_i}\leftarrow c - d_{v_i}\\
				
				\quad 3.~r_{v_i}\leftarrow w+x_id_{v_i} \\			
				\quad 4.~r_{v_i}\leftarrow r_{v_i}mod~{nn}\\			
				\pi\gets (\forall l: \{A_l\},\{B_l\},&&\\
				\quad\quad \quad\{r_l\},\{d_l\})&&\\
				
				&\sends{\pi}
				&\\
				
				& &c\leftarrow \Hzk(\{\{A_l\},\{B_l\}\}_l)\\
				
				& &\sum_{l} d_{l} \stackrel{?}{=} c\\
				
				&& \forall l\in \{1,..,k\}\\

				&&\quad 1.~{r_l}G\stackrel{?}{=}{A_l}+{d_l}X\\
				&&\quad 2.~{r_l}H\stackrel{?}{=}{B_l}-d_lF_l\\
				&&\quad\quad\quad\quad+d_lV_i\\
			\end{protocolm}
		}
	\end{center}
	\caption{ZKP of membership for 1-out-of-$k$ choices on the elliptic curve.
	}
	\label{fig:ECCZKPMultiCandidate}
\end{figure}

\noindent\textbf{Fault Recovery.}
Each pair of participants $P_i$ and $P_j$ ($i\neq j$) have their ephemeral public keys ${x_i}G$ and ${x_j}G$ published on the blockchain.
Hence, $P_i$ using her private key $x_i$ can compute ${x_ix_j}G$ and $P_j$ with her private key $x_j$ can compute  ${x_jx_i}G$. 
Thus, each pair of participants has their counter-party key material shared.
After all honest participants submit the vote repairing transaction, SC verifies submitted data (see \autoref{fig:ZKPECC}), and if no new faulty participants are detected during a round timeout, the stage is shifted to the tally phase, where the votes of all faulty participants are excluded by SC and repaired blinded votes are used for verification of \autoref{eqn:ecctally}.

\noindent\textbf{Tally.} This step computes the votes received for each of the $k$ candidates.
Once the blinded votes from all the participants are verified (see \autoref{fig:ECCZKPMultiCandidate}), the smart contract proceeds to compute
	\begin{equation}
	\label{eqn:ecctally}
	\sum\limits_{i=1}^{n}V_i=
	\sum\limits_{i=1}^{n}{x_iy_iG}+ F=
	{ct_1}{F_1}+{ct_2}{F_2}+...{ct_k}{F_k},
	\end{equation}
	where $ct_1$ to $ct_k$ are the count of votes for candidates 1 to $k$, respectively. 
	Note that $\sum_{i}x_i y_i.G=0$ as per the self-tallying property.
	The individual counts as earlier (see ~\autoref{ssec:basicprotocol}, Phase 4. Tally) are found through an exhaustive search.

\begin{figure}[b]
	\begin{center}
		\fbox{\scriptsize
			\begin{protocolm}{2}	
				
				\participants{\underline{$Participant~P_i$}}{\underline{Smart Contract}}
				\participants{\underline{$A\leftarrow {x_i}G,B
						\leftarrow {x_j}G, x_i$}}{\underline{$A\leftarrow {x_i}G,~B\leftarrow {x_j}G$}}
				Let~w_i, \in_{R} \mathbb{Z}_{nn}\\
				C\leftarrow {x_i x_j}G\\
				m_1\leftarrow {w_i}{G}\\
				m_2\leftarrow {w_i}{B}\\

				c\leftarrow  \Hzk(A,B,m_1,m_2)\\ 	
				r_i\leftarrow w_i+ cx_i &&\\
				&\sends{C,r_i,m_1,m_2} & c\leftarrow  \Hzk(A,B,m_1,m_2)\\

				&  & rG\stackrel{?}{=}{m_1} + cA\\
				&  & rB\stackrel{?}{=}{m_2} + cC\\

			\end{protocolm}
		}
	\end{center}
	\caption{ZKP verifying correspondence of $x_i x_j G$ to PKs $A= x_iG, B=x_jG$.}
	\label{fig:ZKPECC}
\end{figure}
 
\section{Vote Privacy}
\label{sec:appendixvoteorivacy}	
\textit{\textbf{Lemma:}} 
The \name protocol preserves the privacy of blinded votes in all cases other than unanimous voting \textit{iff} there are at least three participants and at least two of them are honest, i.e., $n\ge3 \wedge t\leq n - 2$ 

\noindent
\textit{\textbf{Proof:}} 	
In the case of two participants, their MPC keys (\autoref{eqn:eq1}) are:
 \begin{eqnarray}
 	g^{y_1} &=& g^{-x_2}, \\
 	g^{y_2} &=& g^{x_1},
 \end{eqnarray}
 and their blinding keys are 
 \begin{eqnarray}
 g^{x_1 y_1} &=& g^{-x_1 x_2}, \\
 g^{x_2 y_2} &=& g^{x_1 x_2},
 \end{eqnarray}
 which are inverses to each other, making it trivial to determine the vote choice of another participant by inverting out the blinding key from blinded votes
 \begin{eqnarray}
 B_1 &=& g^{x_1 y_1} f_{p_1} = g^{-x_1 x_2} f_{p_1},\\ 
 B_2 &=& g^{x_2 y_2} f_{p_2} = g^{x_1 x_2} f_{p_2}.
 \end{eqnarray}
 
\noindent 
For $n=3$, each blinding key has at least one part of the key material (required to unblind the vote of the other participants) not available to one participant:
 \begin{eqnarray}
B_1 &=& g^{x_1 y_1} f_{p_1} = g^{-x_1 x_2 - x_1 x_3} f_{p_1},\\ 
B_2 &=& g^{x_2 y_2} f_{p_2} = g^{x_1 x_2 -x_2 x_3} f_{p_2},\\
B_3 &=& g^{x_3 y_3} f_{p_3} = g^{x_1 x_3 + x_2 x_3} f_{p_3}.
\end{eqnarray}
As a result, even if one of the participants were to be dishonest, the vote privacy of the remaining two honest participants would be preserved.

\noindent 
For $n\geq 3$,  participant $P_i$'s blinding key $g^{x_i y_i}$ has its key material $g^{x_i x_j}$ shared only with its counter-party and would require the remaining $n-1$ participants to collude for the recovery of the honest participant's blinding key.
Therefore, the protocol preserves voter privacy if there are at least two honest participants.

\section{Blockchain Voting vs. Centralized Voting}
\label{sec:appendix-blockchain-vs-centralized-voting}	In this section, we compare blockchain  e-voting with centralized e-voting since blockchain e-voting has received some critique~\cite{critic-blockchain-for-voting,park2021going} in the past. 
We use \name as an in-line example and we conclude that blockchain voting can only improve the potential problems of centralized voting at the cost of more complex governance.

\myparagraph{\textbf{Bulletin Board vs. Blockchains}} 
The definition of a bulletin board~\cite{Kiayias2002} assumes its immutability and append-only feature.
While the bulletin board is a conceptual data structure, its expected behavior is not possible to perfectly guarantee neither in centralized voting nor blockchain voting.
However, in centralized voting, a single entity can decide about removing/modifying an entry in the bulleting board, while in blockchain voting such a decision has to be made by a quorum of the honest majority. 
Hence, the blockchain intuitively provides stronger append-only and immutability guarantees.

\myparagraph{\textbf{Device Exploitation at Participants}} 
Park et al.~\cite{park2021going} discuss several categories of attacks on electronic voting systems.
The first of them is \textit{device exploitation}, in which $\mathcal{A}$ tampers with the hardware or software of $P$s, and thus she can change cast vote for the malicious one -- we argue that this attack is the same in both centralized and blockchain voting, and it depends on the protections made at the $P$'s side.
To thwart such attacks, signing of the vote should be done in a dedicated tamper resistance device (i.e., hardware security module (HSM)), which stores the private keys of $P$, while the device must contain a display capable of presenting the data being signed to $P$.
Although such devices might be used in both types of voting, they are more common in the blockchain scenario, where they are referred to as \textit{hardware wallets} (e.g., Trezor, Ledger, KeepKey).
We refer the reader to the work of Arapinis et al.~\cite{ArapinisGKK19} for the security analysis of hardware wallets, where the authors emphasize the importance of ``see-what-you-sign'' on a display of a HW wallet to thwart tampering. 

\myparagraph{\textbf{Device Exploitation at Voting System}}
Another option for a device exploitation is compromising the voting system itself (i.e., its infrastructure).
In the case of centralized voting, the difficulty (and also the cost) of compromising a single centralized e-voting system  is lower than compromising a majority of consensus nodes\footnote{I.e., nodes with a majority of consensus power.} in a decentralized infrastructure of blockchain.
Park et al.~\cite{park2021going} state that device exploitation within e-voting infrastructure may prevent voters from casting their votes as intended (and thus violating E2E verifiability~\cite{benaloh2015end}). 
We argue that while there is no difference in device exploitation of the client between both voting types, centralized voting is significantly more vulnerable to censorship and availability attacks when device exploitation occurs on its centralized infrastructure, as opposed to decentralized blockchain voting that provides extremely high availability (see also below).

\myparagraph{\textbf{(D)DoS Attacks}}
The authority of centralized voting is exposed to a potential (D)DoS attack during the whole period of running.
In contrast, $VA$ in \name is exposed to a potential DoS attack only during the registration phase while the remaining phases run under high availability (depending on a particular blockchain).

\myparagraph{\textbf{Censorship}}
Blockchain voting improves the censorship resistance problem of centralized voting -- intuitively, it is more likely that the vote censorship will be perpetrated due to a decision made by a single entity in centralized voting than the majority of consensus power in the blockchain.

Josh Benaloh went further and put the censorship into the position for the conflict-of-interest~\cite{critic-blockchain-for-voting}: ``\textit{Suppose the transactions are votes, and I am the leader of a movement to oppose a heavy tax on blockchain miners. If I am going to vote in that referendum, then I have to convince some blockchain miner to pick up my vote and put it into the chain.}''
We argue that in the case of suspicion on a conflict-of-interest, the most convenient blockchain platform can be selected for the voting protocol (i.e., \name) since each protocol run is independent.
Moreover, running \name over a public permissionless blockchain might be too expensive for particular use cases, such as large-scale elections. 
For such use cases, more suitable types are permissioned or semi-permissioned~\cite{sra-survey} blockchains running a Proof-of-Stake and/or BFT consensus protocols (see \autoref{ssec:cost-perf-limitations}). 
Nevertheless, in any type of blockchain,  multiple consensus nodes substitute a part of functionalities provided by a single authority in centralized voting, which improves the censorship resistance.\footnote{Note that identity verification and registration can remain centralized, as might be requested by organizers of large-scale elections.}

Finally, a single honest consensus node is sufficient to provide the public with indisputable proof of vote censorship perpetrated by a dishonest majority if it occurs -- such a proof would contain a valid block including a censored transaction, which was produced by an honest node but was abandoned in an orphaned fork by the dishonest majority. 
Furthermore, a hash of such a block might be timestamped by a secure timestamping service to provide proof that the block existed in the time of fork occurrence  (e.g., OpenTimestamps, POEX.IO, STAMPD).
This is not possible in centralized voting since voting authority can contend the voter had not cast her vote.

\myparagraph{\textbf{Lost Private Keys}}
Park et al.~\cite{park2021going} state that $P$s who lose their blockchain keys are unable to vote and recovery of the key in is not possible, as opposed to centralized voting.
We argue that \name (and blockchain voting in general) is a one-time short process ranging from a few hours to a few days during which $P$s generate a new key-pair for interaction with the blockchain as well as a new ephemeral key-pair for homomorphic encryption.
Hence, the probability of losing a newly generated key in such a short time is very low.

\myparagraph{\textbf{Coercion and Vote Selling}}
When comparing centralized voting with block\-chain voting, the coercion and vote-selling problems remain the same -- both are possible due to the remote location of voting, while the blockchain does not bring any disadvantages in contrast to centralized voting.
Besides, the coercion resistance (and thus receipt-freeness), which is in general attributed to local voting systems, is not guaranteed in these systems if trusted $VA$ misbehaves and involves in it~\cite{chaum2008scantegrity}.

\myparagraph{\textbf{Data Feeds}}
Park et al.~\cite{park2021going} emphasize issues of decentralized data feeds (a.k.a., oracles) such as Augur~\cite{peterson2015augur}, ChainLink~\cite{ellisdecentralized}, Witnet~\cite{de2017witnet}.
However, we argue that the blockchain voting does not require any data feed service and can run on a smart contract platform using solely the data submitted by $VA$ and $P$s.

\myparagraph{\textbf{Fixing Bugs}}
Park et al.~\cite{park2021going} criticize additional complexity and required coordination related to fixing bugs across the infrastructure of blockchains in contrast to centralized voting. 
However, this is considered a tax for decentralization and other beneficial blockchain features. 
Nevertheless, such coordination is manageable in all blockchain types and is usually defined in the governance model. 
\end{document}